\documentclass[aps,prd,reprint,superscriptaddress,nofootinbib,floatfix]{revtex4-2}
\usepackage{ifxetex}

\ifxetex
   \usepackage{polyglossia}
   \setdefaultlanguage{english}
   \usepackage{fontspec}
\else
   \usepackage[utf8]{inputenc}
\fi

\usepackage{amsmath}
\usepackage{graphicx,color,overpic,mathtools}
\usepackage{amsthm}
\usepackage{bm}
\usepackage{amsmath}
\usepackage{upgreek}
\usepackage{amssymb}
\usepackage{setspace}
\usepackage{hyperref}
\usepackage{xcolor}
\PassOptionsToPackage{normalem}{ulem}
\usepackage{ulem}
\usepackage{tensor}

\usepackage{graphicx,color}
\usepackage{bbold}
\usepackage{overpic}

\usepackage{pgf}
\usepackage{tikz}

\newcommand{\noprint}[1]{}

\renewcommand{\d}{\mathrm{d}}

\newcommand{\caslavcom}[1]

\usepackage{bbm}

\usepackage{bbm}

\usepackage{amsmath,amsfonts,amssymb,amsthm,amsbsy,mathtools}
\usepackage{graphicx}
\usepackage{bm}
\usepackage{physics} 
\usepackage{bbold}
\usepackage{mathtools}
\usepackage{xcolor}
\usepackage{tikz}
\usepackage{tikz-cd}
\usepackage{comment}

\usepackage{comment}
\usepackage{soul}
\usepackage{framed}
\usepackage{mdframed}
\usepackage{csquotes}
\usepackage{appendix}
\usepackage{hyperref}

\usepackage{natbib}
\usepackage[normalem]{ulem}
\usepackage{xcolor}
\usepackage{xspace}
\usepackage{ifthen}
\usepackage{braket}
\usepackage[dvipsnames]{xcolor}
\usepackage{hyperref}

\renewcommand{\d}{\mathrm{d}}
\graphicspath{ {./images/} }
\DeclareMathOperator{\supp}{supp}

\newcommand\notbackslash[1]{%
  \mathrel%
  {%
    \ooalign{\hidewidth$\backslash$\hidewidth\cr$#1$}%
  }%
}

\begin{document}

\title{
Factorisation conditions and causality for local measurements in QFT}

\begin{abstract}
    Quantum operations that are perfectly admissible in non-relativistic quantum theory can enable signalling between spacelike separated regions when naively imported into quantum field theory (QFT). Prominent examples of such “impossible measurements” include certain unitary kicks and projective measurements. It is generally accepted that only those quantum operations whose physical implementation arises from a fully relativistically covariant interaction, between the quantum field and a suitable probe, should be regarded as admissible. While this idea has been realised at the level of abstract algebraic QFT, or via particular measurement models, there is still no general set of operational criteria characterising which measurements are physically implementable. In this work we adopt the local $S$-matrix formalism to describe local operations, and make use of a hierarchy of factorisation conditions that exclude both superluminal signalling and retrocausality, thereby providing such a criterion. Realising the local $S$-matrices through explicit interactions between smeared field operators and a pointer degree of freedom, we further derive local causality conditions for the induced Kraus operators, which guarantee the absence of signalling in “impossible measurement” scenarios. Finally, we show that the accuracy with which local field observables can be measured is fundamentally limited by the retarded propagator of the field, which plays an essential role in the factorisation identity we prove for the field Kraus operators. 
    \end{abstract}


\author{Robin Simmons}
\affiliation{University of Vienna, Faculty of Physics, Vienna Doctoral School in Physics, and Vienna Center for Quantum Science and Technology (VCQ), Boltzmanngasse 5, A-1090 Vienna, Austria}
\affiliation{Institute for Quantum Optics and Quantum Information (IQOQI),
Austrian Academy of Sciences, Boltzmanngasse 3, A-1090 Vienna, Austria}

\author{Maria Papageorgiou}
\affiliation{Institute for Quantum Optics and Quantum Information (IQOQI),
Austrian Academy of Sciences, Boltzmanngasse 3, A-1090 Vienna, Austria}

\author{Marios Christodoulou}
\affiliation{Institute for Quantum Optics and Quantum Information (IQOQI),
Austrian Academy of Sciences, Boltzmanngasse 3, A-1090 Vienna, Austria}

\author{\v{C}aslav Brukner}
\affiliation{University of Vienna, Faculty of Physics, Vienna Doctoral School in Physics, and Vienna Center for Quantum Science and Technology (VCQ), Boltzmanngasse 5, A-1090 Vienna, Austria}
\affiliation{Institute for Quantum Optics and Quantum Information (IQOQI),
Austrian Academy of Sciences, Boltzmanngasse 3, A-1090 Vienna, Austria}

\date{\small\today}

\maketitle

\section{Introduction}
The topic of impossible measurements \cite{sorkin1993impossible} has received renewed attention over the past few years. A multitude of approaches have been developed to address the impossible measurement scenarios (see \cite{papageorgiou2024eliminating} and references therein). The clear moral that has emerged is that the set of quantum mechanical operations, or channels, typically allowed in quantum information is far too large when brought to bear on relativistic theories. In the other direction, when one formulates local measurement theory for relativistic quantum field theory (QFT), the resulting channels do not necessarily include the most familiar ones from quantum information. 

A straightforward implication is that the quantum informational machinery of states and operations cannot be straightforwardly imported into QFT. In fact, the established empirical success of QFT relies on the asymptotic scattering paradigm, which is not suitable for describing local operations, i.e. operations over bounded regions of spacetime (for historical context see \cite{Fraser2023}). Sorkin's examples of impossible operations \cite{sorkin1993impossible}, and recent refinements \cite{PhysRevD.104.025012}, showed that operations built out of local operators (i.e. local Kraus operators) generally allow superluminal signalling. Specifically, this includes ideal measurements of many important observables \cite{Benincasa_2014, albertini2023ideal, PhysRevD.105.025003}, leading to a number of papers suggesting non-ideal measurement channels, both abstract \cite{PhysRevD.105.025003, Gisin2024towardsmeasurement, albertini2023ideal, oeckl2024spectraldecompositionfieldoperators} and concrete i.e. derived from some purified model \cite{ 
mandrysch2024quantumfieldmeasurementsfewsterverch}. The former derive a number of no-signalling conditions on the abstract channels or their Kraus operators, while the latter rely on properties of the purified dynamics.

From the opposite direction, a rigorous framework for quantum measurement theory in algebraic QFT has been proposed by Fewster and Verch (FV framework)\cite{fewster2020quantum, PhysRevD.103.025017,fewster_asymptotic_2023}, based on purified dynamics. Together, these results point towards a general theory of measurement and operations in QFT. However, there is still no general set of operational criteria, in terms of e.g. Kraus operators or their purifications, that characterises which measurement maps and operations are physically implementable. A fundamental step towards this would be a ``causal Stinespring theorem'', an idea on which we comment at the end of this paper.  

The main results of the paper are as follows. We adapt the local $S$-matrix formalism due to Bogoliubov \cite{bogoliubov1960introduction, blum2025perturbative} to describe local operations, and we analyse the setting of Sorkin's impossible operations. 
A key result is the sufficiency of a simple factorisation condition, which allows the decomposition of a local operation induced by the local $S$-matrix $S = S[f]$, where $f$ is a compactly supported test function. This decomposition is expressed as:
     \begin{equation} 
        S[f]= S[f_+]S[f_-] \label{Sf_pm}
    \end{equation}
where $f=f_++f_-$, splitting the support of $f$ into past ($f_-$) and future ($ f_+$) components with respect to any spacelike hypersurface passing through the support of $f$. We refer to this property as the \textit{continuous additivity} of the local $S$-matrix. However, this property may not hold for certain local interactions. For this reason, we also demonstrate the sufficiency of a weaker factorisation condition, known as the \textit{Hammerstein property} of the local scattering maps. The Hammerstein property plays a central role in capturing the locality of interactions within the causal perturbation theory framework, which formalises renormalization \cite{Epstein:1973gw}, perturbative algebraic QFT \cite{Rejzner:2016hdj}, as well a recent approach to a non-perturbative definition of the local $S$-matrix in quantum field theory \cite{Buchholz_2020}.

    Further, when operations are induced  by the local coupling of the field to a probe system, we examine the consequences of these factorisation conditions for the induced operations on the field's Hilbert space after tracing out the probe. Such operations are defined by means of Kraus operators $\{\Pi^{\psi}_q[f]\}$ where $\psi,q$ denote states in the probe's Hilbert space. We show that continuous additivity when combined with the assumption that the $S$-matrix implements quantum control, leads to a no-signalling condition on the associated Kraus operators,
    \begin{equation} \label{Robert}
        \Pi_q^{\psi}[f]=\int \d s\, \Pi_{q-s}^{\psi_+}[f_+]\Pi_s^{\psi_-}[f_-].
    \end{equation}
      
      Finally, we use an exactly solvable concrete measurement model that can be seen as a relativistic von Neumann measurement model, where a local pointer variable is linearly coupled to a smeared scalar field operator via $\mathcal{L}_I[f]=\phi(f)\otimes P$ \cite{PhysRevD.109.065024}. We find that the $S$-matrices and Kraus operators obey the conditions \eqref{Sf_pm} and \eqref{Robert}, and so this forms a concrete example of the link between the two representations of channels and no-signalling conditions. Further, we obtain an explicit class of no-signalling measurement channels on the field whose sharpness is fundamentally limited by the retarded field propagation within the support of $f$, thereby imposing an intrinsic bound on the accuracy of local field measurements. These channels coincide with the abstract continuous measurement channels studied in \cite{PhysRevD.105.025003, albertini2023ideal, oeckl2024spectraldecompositionfieldoperators}, once this new sharpness constraint is taken into account. 

The paper is structured as follows: In section \ref{sec:ops_and_measurements_QFT} we briefly introduce standard measurement theory and a minimal (algebraic) QFT set-up. We review an abstract notion of operation in the QFT set-up, and briefly present the challenges posed by the impossible measurements scenario, using a simple example of impossible unitary operations introduced in \cite{PhysRevD.105.025003}.  In section \ref{subsec:rel_dynamics} we introduce the local $S$-matrix formalism, to discuss the properties of relativistically valid unitary operations in QFT that can in principle represent local (not asymptotic) interactions. We take the local $S$-matrix to represent the local coupling of the field to a probe system \cite{hellwigkraus1970b} and we examine the operational consequences of a hierarchy of factorisation properties of the $S$-matrix. 

In section \ref{sec:Operations induced by probes} we explicitly introduce field operations induced by probes, and the induced Kraus operators in the field's Hilbert space that are fixed by the local $S$-matrix and the characteristics of the probe system. We consider a simple model for measurements of smeared field amplitudes $\phi(f)$, and consider the limitations to the accuracy of field measurements that come from the causal (retarded) field propagation within  the support of the test function $f$. Specifically, the sharpness of a measurement of $\phi(f)$ is bounded from below by $|\Delta_{\textsc{r}}(f,f)|^{1/2}$, i.e. the square-root of the retarded Green's function smeared against $f$. We show how this model relates to the families of Kraus operators for field measurements introduced  in Albertini and Jubb \cite{albertini2023ideal} and Oeckl \cite{oeckl2024spectraldecompositionfieldoperators}.  

In section \ref{subsec:causality_for_example} we derive no-signalling conditions of the induced Kraus operators that can ensure no superluminal signalling and no retrocausality. We discuss the necessity or sufficiency of these conditions and for what type of models they are expected to hold. Finally, we discuss the implications for von Neumann-type measurements of more general QFT observables and how this analysis provides hints for relativistic dilation theorems in QFT. Note that the main results are in sections \ref{subsec:rel_dynamics} and \ref{subsec:causality_for_example}, while long calculations and proofs are in the appendices.

\section{Local operations in QFT}\label{sec:ops_and_measurements_QFT}

Perhaps surprisingly, local measurement theory for QFT is still being developed today. This is because the standard predictions and experiments of QFT take the form of asymptotic scattering amplitudes, yet this is not sufficient for modelling local experiments in a relativistic setting. The main obstacle in extracting local probabilistic predictions from QFT theory is that standard quantum mechanical prescriptions for  ideal, or projective, measurements and unitary kicks cannot be straightforwardly applied. These operations, when extended to QFT, can lead to signalling across spacelike separated regions. The problem of impossible operations, which was first introduced by Sorkin \cite{sorkin1993impossible}, was meant to highlight the friction between no-signalling and the quantum operations that are typically assumed in quantum mechanics. For examples see \cite{Benincasa_2014} and more recently \cite{PhysRevD.105.025003, albertini2023ideal}.

Quantum mechanics has a standard measurement theory, see \cite{busch_book_2016, busch_standard_1996}. Despite ongoing ambiguity regarding what precisely qualifies certain interactions as measurements -- a central aspect of the quantum measurement problem -- there exist standard prescriptions for extracting probabilistic predictions for any measurement or general operation defined within the theory. In this section we briefly review the basics of standard quantum measurement theory, we introduce a minimal (algebraic) framework for QFT and we discuss the problem of `impossible measurements'. We also briefly review the Kraus and Stinespring representations of channels, which will appear in our analysis of QFT channels.

\subsection{Standard measurement theory}

A projective measurement of an observable \( A \), represented as a (compact) self-adjoint operator acting on a Hilbert space $\mathcal{H}$, with spectral decomposition
\begin{equation}
A = \sum_j a_j P_j,
\end{equation}
involves a set of orthogonal projectors \( P_j \) corresponding to the possible outcomes \( a_j \). If the system is initially in a state \( \rho \), and the outcome \( a_j \) is obtained, the post-measurement state is given by the Lüders' rule in its selective form:
\begin{equation}
\rho \mapsto \frac{P_j \rho P_j}{\text{Tr}(P_j \rho)},
\end{equation}
where the denominator is the probability of that outcome occurring.
If, however, the measurement is performed without recording which outcome occurred---a non-selective measurement---the state updates according to the trace-preserving map
\begin{equation} \label{NS}
\rho \mapsto \sum_j P_j \rho P_j,
\end{equation}
which we call an ideal non-selective update. A projective measurement is called ideal, or repeatable. 

An observable defines a family of projection operators, called a projection valued measure (PVM), through its spectral decomposition. For an ideal measurement, the elements of the PVM directly relate the probabilities of an outcome, by
\begin{equation}
    p_i(\rho)=\text{Tr}(P_i\rho).
\end{equation}
Here we can see the repeatability. The probability of getting $i$ a second time, given the outcome of the first measurement was $i$ is given by
\begin{align}
    p_{i|i}(\rho):=&p_i(\rho_i)=\text{Tr}(P_i\rho_i)\nonumber \\=&\frac{\text{Tr}(P_iP_i  \rho P_i)}{\text{Tr}(P_i \rho)}=\frac{\text{Tr}(P_i^3 \rho )}{\text{Tr}(P_i \rho)}=1,
\end{align}
where we have used the cyclicity of the trace to move all projections $P_i$ together, and then used that $P_i^2=P_i$. 

A non-ideal, i.e. non repeatable, measurement must be described by a set of positive operators $E_i$ that do not form a complete orthogonal set of projections, but do still compute the probabilities of different outcomes,
\begin{equation} \label{prob}
    p_i=\text{Tr}(E_i\rho).
\end{equation}
Such a family, which sums to the identity, is called a positive operator valued measure (POVM). 

The non-selective update defined in Eq \eqref{NS} corresponds to a completely positive trace-preserving (CPTP) map acting on the system. More generally, any such CPTP map can be represented as arising from a unitary interaction between the system and an ancillary probe (or environment), followed by tracing out the probe. That is, for every CPTP operation \( \tilde\Phi \) on the system, there exists a unitary \( U \) acting on the joint system--probe space and an initial probe state \( \sigma \) such that
\begin{equation}  \label{channel}
\tilde{\Phi}(\rho)= \text{Tr}_{\text{probe}}\big[ 
U (\rho \otimes \sigma) U^\dagger \big],
\end{equation}
where $\tilde{\Phi}(\rho)$ is the Schrödinger picture channel, and this representation is given by Stinespring's theorem. This `dilated' picture provides a powerful and physically grounded way to understand general CPTP quantum operations as arising from purified dynamics on an extended Hilbert space. The Kraus representation theorem states that the channel \eqref{channel} can be represented by a set of Kraus operators $\{\Pi_i\}$ on $\mathcal{B}(\mathcal{H})$, where $i$ is indexing an orthonormal set in the probe's Hilbert space, and such that $\Pi^{\dagger}_i \Pi_i := E_i$ defines a POVM. Consider measuring the probe ideally after the evolution, and finding it in state $\ket{i}$, then the updated state is 
\begin{equation}
    \rho_i=\frac{\Pi_i\rho \Pi^{\dagger}_i}{p_i},
\end{equation}
where $p_i$ is given by Eq \eqref{prob}. Summing over all possible probe outcomes, the non-selective update map becomes 
\begin{equation}
    \tilde{\Phi}(\rho)=\sum_ip_i\rho_i=\sum_i \Pi_i\rho \Pi_i^\dagger, \; 
\end{equation}
which is equivalent to Eq \eqref{channel} by the Kraus representation theorem. $\tilde{\Phi}$ is a CPTP map thanks to the fact that the Kraus operators satisfy 
\begin{equation}
    \sum_i \Pi_i^\dagger \Pi_i= \openone.
\end{equation}
We can also define the channel $\Phi$ in the Heisenberg picture, acting on operators rather than states
\begin{equation}
      \Phi(O)=\sum_i \Pi^\dagger_i O \Pi_i, \; 
\end{equation}
which is dual to $\tilde{\Phi}$ in the sense 
\begin{equation}
    \text{Tr}(\rho \Phi(O))=\text{Tr}(\tilde\Phi(\rho)O),
\end{equation}
for all states $\rho$ and operators $O$. $\Phi$ is a unital channel, i.e. a channel that maps the identity operator to itself $\Phi(\openone)=\openone$.
In QFT it is natural to use the (dual) Heisenberg picture of the channels. Note that in this paper we will mostly consider Kraus-representable \textit{non-selective}, i.e., UCP channels on the field.

\subsection{Set-up}\label{subsec:sorkin}

 Below we introduce the QFT set-up. We focus on the real scalar field and we introduce the notion of a local operation, which is key for our understanding of Sorkin's impossible operations. Finally, we review an example from \cite{PhysRevD.105.025003} that is most relevant for the measurement model that we will consider in section \ref{sec:example}.

Consider a canonically quantised real scalar field theory on Minkowski spacetime $\mathcal{M}$. The field operator $\phi(\mathsf x)$ at a point $\mathsf x \in \mathcal{M}$ is an operator valued distribution and so in order to form an operator algebra we define smeared field operators
\begin{equation}
    \phi(f)=\int_\mathcal{M} \text{dVol } f(\mathsf x) \phi(\mathsf x) ,
\end{equation}
for any $f\in C^\infty_0(\mathcal{M})$, which is the set of all smooth, compactly supported test functions on $\mathcal{M}$. We further define
\begin{align}
    [\phi(f),\phi(g)]&=i\int_{\mathcal{M}^2} \text{d}^2\text{Vol } \Delta(\mathsf x,\mathsf y)f(\mathsf x)g(\mathsf y) \openone \nonumber \\
    &:=i\Delta(f,g) \openone, \label{generators}
\end{align}
where $\Delta(\mathsf x,\mathsf y)$ is the usual Pauli-Jordan form
\begin{equation}
   \Delta (\mathsf x,\mathsf y)= \Delta_{\textsc{a}} (\mathsf x,\mathsf y)-\Delta_{\textsc{r}} (\mathsf x,\mathsf y)  
\end{equation}
where $\Delta_{\textsc{a},\textsc{r}}$ the advanced and the retarded Green's functions respectively. Note that $\Delta(f,g)=0$ if $\supp f, \supp g$ are spacelike. The commutation relation Eq \eqref{generators} can be viewed as a condition that must be satisfied by the generators of the local algebras. 

In this real scalar field model, abstract local algebras $\mathcal{A}(\text{R})$ can be formed by finite degree polynomials of $\openone$ and $\phi(f)$ for $\supp{f}\subset \text{R}$. When working in Minkowski spacetime, we can pick the vacuum state $\omega:\mathcal{A}(\mathcal{M})\rightarrow \mathbb{C}$ and construct the Gelfand–Naimark–Segal (GNS) representation $\pi:\mathcal{A}(\mathcal{M})\rightarrow \mathcal{L}(\mathcal{H})$. From now on we will consider all operators to be acting on this Hilbert space, and drop $\pi$. The vacuum vector state $\ket{0}\in \mathcal H$ is contained in a dense domain $D_0$ on which all elements of the field algebra are well-defined. 

The Weyl operators $W(f)=e^{i\overline{\phi(f)}}$ form a representation of the canonical commutation algebra,
\begin{equation}
    W(f)W(g)=W(f+g)e^{-i\Delta(f,g)/2}.
\end{equation}
Taking the Weyl operators with $\supp{f}\subset$ R as generators we can construct a Weyl algebra for each region R, denoted $\mathcal{W}(\text{R})$. Finally, we can take the double commutant $\mathfrak{A}(\text{R}):=\mathcal{W}(\text{R})''$---where the commutant of $S\subseteq \mathcal{B}(\mathcal{H})$ is the set of all operators which commute with all elements of $S$–––which associates a concrete von Neumann algebra to every region R. We consider an operator $A$ to be localised\footnote{$A$ is localised to R if $A$ is affiliated to $\mathfrak{A}(\text{R})$. } to R if $A$ commutes with all operators spacelike to R, i.e. $[A, \mathfrak{A}(\text{R}')]=0$ where $\text{R}'$ is the set of all points spacelike to R. An example is $\phi(f)$, which is local to $\supp{f}\subset \text{R}$, despite being unbounded and thus not in $\mathfrak{A}(\text{R})$. It follows that any bounded function of $\phi(f)$, defined by the spectral calculus, is in $\mathfrak{A}(\text{R})$. For more details on this construction, see \cite{fewster2019algebraic}. Using the properties of the commutator, it follows, from strong commutativity of the fields \cite{Borchers:1992rq}, that spacelike algebras commute,
\begin{equation}
    [\mathfrak{A}(\text{R}), \mathfrak{A}(\text{R}')]=0,
\end{equation}
where regions $\text{R}$ and $\text{R}'$ are space-like separated. More generally, in axiomatic approaches to QFT, e.g. algebraic QFT (AQFT)\cite{haag_kastler_1964}), one postulates the axiom of \textit{microcausality}, namely that spacelike separated algebras of observables commute. This is a kinematical `separability' assumption that, roughly, splits the field degrees of freedom between causally disconnected parts \cite{earman2014relativistic}. Note that microcausality alone does not have an a priori operational consequence \cite{Ruetsche2021, calderon2024causal, soulas2025proof} e.g. in terms of signalling. Once a notion of operation is introduced, microcausality alone is generally not sufficient for blocking superluminal signalling (see also \cite{redei_how_2010}). Sorkin \cite{sorkin1993impossible} showed that, despite the microcausality assumption, even the most basic local operations in QFT, such as unitary kicks and ideal measurements
-- when followed by the Lüders post-measurement state update for non-selective measurements -- can lead to superluminal signalling between spacelike separated regions (see also \cite{PhysRevD.104.025012, PhysRevD.105.025003,oeckl2024spectraldecompositionfieldoperators}).

Let us first introduce an abstract notion of local operation in QFT, as opposed to local observables. By local observables we mean the self adjoint elements of the local algebras $A\in \mathfrak{A}(\text{R})$ as defined above. That is, field observables are represented by self-adjoint operators as in usual quantum theory, but explicitly associated to spacetime regions. We define a local operation to be any completely positive map over the elements of the local algebra\footnote{Quantum channels are usually defined on algebras of bounded operators, as bounded operators do not have domain problems. The dual of such a channel is a CPTP map on density operators. In QFT we will be applying the channels to potentially unbounded operators, like $\phi(g)$. In general there will be domain issues, however the form of channels that we explicitly consider are Gaussian or Bosonic channels, and so they map the field algebra to itself, see \cite[12.4.2]{holevo2019quantum}. } such that $\Phi(\openone)=\openone$ and such that 
\begin{equation}
    \Phi|_{\mathfrak{A}(\text{R}')}=\text{id} \label{localop}
\end{equation}
the identity map, for all regions R$'$ spacelike to R (see also \cite{PhysRevD.105.025003}). This formalises the intuition that a local operation, that in principle is performable within the region R, should not have an effect on regions spacelike to R. This includes unitary kicks, e.g. $e^{iA}$, or \textit{non-selective} ideal measurements of $A$. Note that selective operations, which we are excluding here, are not local in the sense of Eq \eqref{localop} \footnote{Note that Eq \eqref{localop} is not satisfied for \textit{selective} operations due to the existence of spacelike correlations in generic QFT states. This is a case that we are not considering, since we are concerned with the signalling properties of trace-preserving operations.}, but typically this is not considered problematic as long as they cannot enable superluminal signalling.

Below, we will briefly review the issue of impossible operations in QFT, i.e., operations that can enable superluminal signalling. This issue highlights that in QFT there is no obvious link between local observables and local operations, as defined abstractly above.

\subsection{Impossible operations}

\begin{figure}[ht]
\centering
\includegraphics[width=1\linewidth]{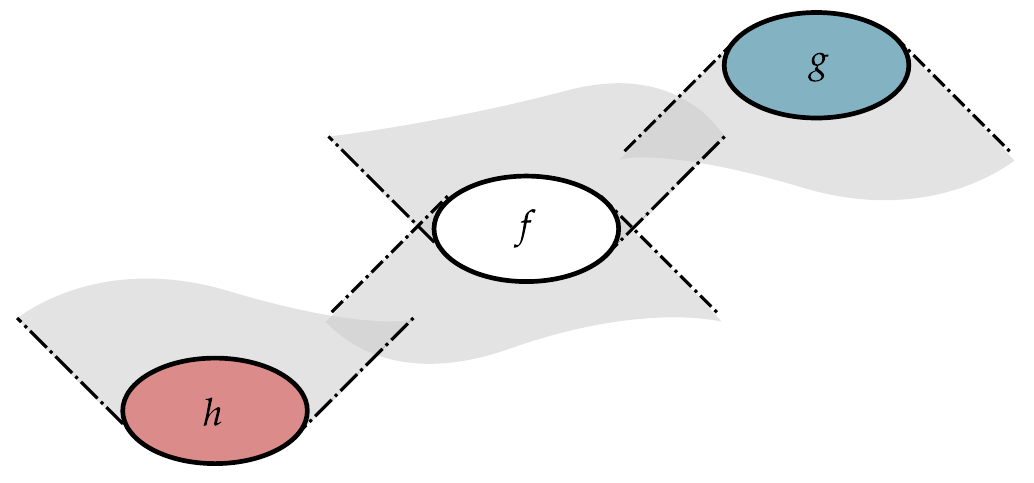}
\caption{Bob's lab (white) is partly in the future of Alice's (red) and partly in the past of Charlie's (blue), while Alice and Charlie are strictly spacelike. Relativistic causality then requires that no signals can be sent from Alice to Charlie, i.e. there must be a no-signalling condition on the experimental setup. Here we have labelled the labs by the supports of the functions $f,g,h$, but the overall causal structure is common to all Sorkin scenarios.}
\label{fig:sorkin}
\end{figure}

Having introduced a minimal AQFT set-up and defined an abstract notion of local operations, we now focus on formalising and presenting Sorkin's impossible operations. This problem highlights that a local operation, i.e. an operation over a local spacetime region, need not be non-signalling, something that can be seen more clearly in a Sorkin-type scenario involving three (or more) regions. This appears in contrast to microcausality, which only refers to pairs of regions. 
In the following we will adopt the notation that regions (singled out by the support of smearing functions $f,g,h$ see figure \ref{fig:sorkin}) will be denoted as A,B,C etc., while local observables will be denoted as $A,B,C$ etc, e.g. $A\in \mathfrak{A}(\text{A})$,  where $\mathfrak{A}(\text{A})$ denotes local algebra, introduced in the next section.

The set-up of the Sorkin problem is the following: consider three regions $\text{A}, \text{B}, \text{C}$ (figure \ref{fig:sorkin}) singled out by the supports of the smearing functions $h,f,g$ respectively. The regions are such that with respect to suitable partial ordering $\text{A}\prec \text{B} \prec \text{C}$ (since B is partially in the causal future of A and C is partially in the causal future of B and the partial order is not transitive), but A and C are spacelike separated.  Two operations $\Phi_{\textsc{{a,b}}}$ are applied over regions A and B. For region A to not signal to region C, the expectation values of observables in C must not depend on the \textit{choice} $\lambda$ of which operation $\Phi^\lambda_{\textsc{a}}$ is implemented in A. Namely, the condition for no impossible operations (which has previously been considered in \cite{PhysRevD.104.025012,deramon2021relativistic, PhysRevD.105.025003, oeckl2024spectraldecompositionfieldoperators}) is simply the following: let $\rho$ be the global state of the field and $C$ an observable over region $\text{C}$, then it should hold that 
\begin{equation} \label{transparency1}
\text{Tr}\left(\Phi_\textsc{a}^\lambda(\Phi_\textsc{b}(C))\rho\right)=\text{Tr}\left(\Phi_\textsc{b}(C)\rho\right), \mbox{ } \forall \lambda.
\end{equation}
Hence, the expectation values of $C$ should not depend on the operation $\Phi_\textsc{a}^\lambda$ that is implemented over the spacelike separated region $\text{A}$. We explicitly introduce a dependence of the operation $\Phi_\textsc{a}$ over some parameter $\lambda$, which could be used to signal to region $\text{C}$ if the no-signalling condition~\eqref{transparency1} fails (see example below).  If one demands this for all states $\rho$ and for all observables $C$, this condition can be viewed as a condition on $\Phi_\textsc{b}$ given $\Phi^\lambda_{\textsc{a}}$.

\subsubsection{Example of signalling  and non-signalling unitaries}\label{subsec:sig_and_non}

The most basic example of failure of this condition using unitary operators in QFT was given in \cite{PhysRevD.105.025003} and uses the real scalar field construction given above. Letting 
\begin{align} \label{unitaries}
&\Phi^{\lambda}_\textsc{a}(\cdot)=e^{i\phi(h_\lambda)}\cdot e^{-i\phi(h_\lambda)}, \nonumber \\
&\Phi_\textsc{b}(\cdot)=e^{i\phi(f)^2}\cdot e^{-i\phi(f)^2}, \nonumber \\
&C=\phi(g)
\end{align}
with supports as shown in figure \ref{fig:sorkin}, application of Baker–Campbell–Hausdorff formula leads to 
\begin{equation}\label{eq:first_sorkin}
\text{Tr}\left(\Phi_\textsc{a}^\lambda(\Phi_\textsc{b}(C))\rho\right)=2\Delta(h_\lambda,f)\Delta(f,g).
\end{equation}
The dependence on $\lambda$ means that Alice can signal to Charlie by picking a family of $h_\lambda$ such that $\Delta(h_\lambda,f)$ is a non-trivial function of $\lambda$. If we replaced  $\Phi_\textsc{b}$ by
\begin{equation}
    \Phi'_\textsc{b}(\cdot)=e^{i\phi(f)}(\cdot) e^{-i\phi(f)}
\end{equation}
then 
\begin{equation}\label{eq:second_sorkin}
\text{Tr}\left(\Phi_\textsc{a}^\lambda(\Phi'_\textsc{b}(C))\rho\right)=\Delta(f,g),
\end{equation}
so there is no $\lambda$-dependence, i.e., no signalling from Alice to Charlie in that case. 
Note that the right-hand side of Eqs. \eqref{eq:first_sorkin} and \eqref{eq:second_sorkin} is independent of the state $\rho$ since it is assumed that $\rho$ is such that $\text{Tr}(C \rho)=0$ (which holds for $C=\phi(g)$ and $\rho$ being e.g. a Fock or thermal state). 

The failure of the no-signalling condition \eqref{transparency1} in Eq \eqref{eq:first_sorkin} can be traced back to the fact that $\Phi_\textsc{b}(C)$ will not commute with $A$ in general, even though $[C,A]=0$ by microcausality \cite{PhysRevD.104.025012}. This is not the case for $\Phi'_\textsc{b}$, since it is easy to check that $[\Phi'_\textsc{b}(C),A]=0$. Then, it is natural to suspect that the reason why $\Phi_{\textsc{b}}(\cdot)= e^{i\phi(f)^2}\cdot e^{-i\phi(f)^2}$ is problematic is that $\phi(f)^2$ is a \textit{non-local operator} (given by a double integration over spacetime in this case). Further, one can speculate that the reason why $\Phi'_{\textsc{b}}(\cdot)= e^{i\phi(f)}\cdot e^{-i\phi(f)}$ is not problematic is the following property of Weyl operators 
\begin{equation}
e^{i\phi(f+f')}=e^{i\phi(f)}e^{i\phi(f')}e^{i\Delta(f,f')/2}
\end{equation}
i.e. that we can split the unitary up to a phase, something that does not hold for the squared field unitary. Intuitively, this is because it implies that $\Phi'_{\textsc{b}}$ can be split into causally ordered parts, which can be chosen so that the part in the future of region $\text{A}$ drops out the relevant expectation values. As we shall see in section \ref{subsec:rel_dynamics} when this type of factorisation property holds for a class of unitaries it indeed rules out impossible operations.

\section{Local \texorpdfstring{$S$}{S}-matrix formalism}\label{subsec:rel_dynamics}
In this section we will consider a broad set of valid local unitaries in QFT, and their properties. These local unitaries can be interpreted as local operations, and we find that there is a natural causal condition on them that ensures they define non-signalling channels. Bogoliubov's local $S$-matrix formalism \cite{bogoliubov1960introduction} is a local and covariant approach to describing interactions in QFT\footnote{The first attempt to formulate relativistic quantum theory in terms of the $S$-matrix is due to Heisenberg, who introduced a `global' $S$-matrix satisfying unitarity and Lorentz invariance. The local $S$-matrix approach was later introduced by Stueckelberg and Bogoliubov, reintroducing local degrees of freedom and demanding additional causality conditions \cite{blum2025perturbative}.}, and so is a natural place for us to start. It plays a major role in perturbative algebraic QFT, which provides a rigorous definition of observables in interacting field theory via a generating function, where the factorisation properties play a key role in proving the main theorem of renormalisation (see \cite{rejzner2019localitycausalityperturbativeaqft} and references within, as well as the complete introduction \cite{Rejzner:2016hdj}).

A local $S$-matrix will be denoted as $S[f]$ where $f$ a compactly supported (smooth) real-valued smearing function. In some cases, when the adiabatic limit can be taken $f(\mathsf x)\rightarrow f_\text{const}$, we end up with the usual global (asymptotic) $S$-matrix familiar from textbook QFT, where coupling terms are spacetime independent constants e.g. $\mathcal{L}_I(\mathsf x)=-f_\text{const} \phi(\mathsf x)^4$, and which map between the in and out Hilbert spaces. While the local $S$-matrix formalism was originally conceived, and 
still used in perturbative AQFT (pAQFT), to define scattering interactions in cases where 
the asymptotic $S$-matrix might not be well defined (for historical context see \cite{fraser2024perturbative,blum2025perturbative}), it also provides
a useful model for more general localised interactions, which may include scenarios beyond scattering. 
Further, as we will see, it comes endowed
with a hierarchy of causal conditions which are very well suited 
for studying causality in finite time processes. For these reasons, we will be concerned only with the local case, where the support of $f$ is bounded.

\begin{figure}[ht!]
    \centering
    \includegraphics[width=0.8\linewidth]{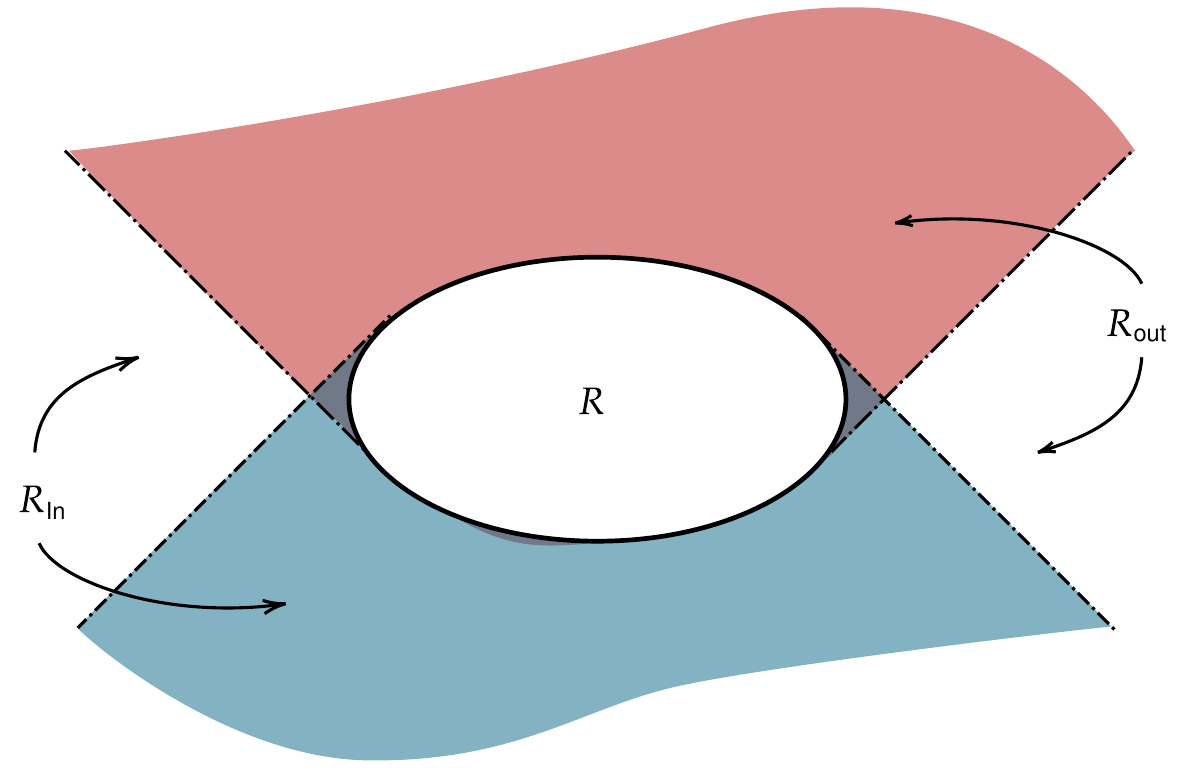}
    \caption{The region $\text{R}_\text{in}=\mathcal{M}\backslash J^+(\text{R})\; (\text{R}_\text{out}=\mathcal{M}\backslash J^-(\text{R}))$ (where $\mathcal{M}$ is the spacetime, and $J^\pm(\text{R})$ refer to the causal future and past of R, the red and blue regions respectively) for a region R, given by the spacelike white regions and the blue (red) region. The out region is the set of all points $\mathsf x$ for which $\mathsf x\not\preccurlyeq \text{\text{R}}$, and likewise the in region is the set of all points $\mathsf y$ such that $\text{R}\not\preccurlyeq \mathsf y$.}
    \label{fig:out_region}
\end{figure}

\begin{figure}
    \centering
    \includegraphics[width=1\linewidth]{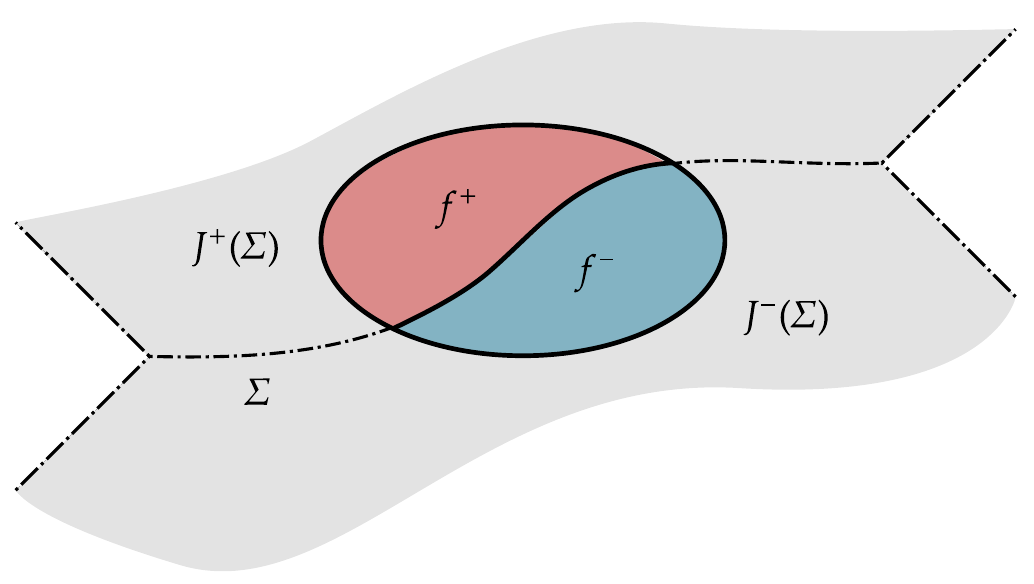}
    \caption{Continuous additivity of an $S$-matrix corresponds to selection of a (partial) Cauchy surface $\Sigma$ that bisects the support of $f$. The grey region above (below) $\Sigma$ is $J^+(\Sigma), \; (J^-(\Sigma))$.}
    \label{fig:split}
\end{figure}

\subsection{Properties of valid local \texorpdfstring{$S$}{S}-matrices}\label{subsubsec:properties_of_smatrices}

Note that we will not be assuming any (formal) expansion for $S$, e.g., the Dyson expansion. Instead, drawing inspiration from a recent non-perturbative approach to the local $S$-matrix \cite{Buchholz_2020}, we will be assuming certain properties of the $S$-matrix that are expected to hold non-pertubatively. We assume that $S=S[f]$ exists for a compactly supported test function $f$, and is a unitary element of the local algebra $\mathfrak{A}(\text{R})$ where $\text{R}=\text{supp}f$, \footnote{Here we assume that $S(f)$ is an element of the local algebra, namely that it is constructed out of operators local to the support of $f$, while in \cite{Buchholz_2020} the local algebras are generated by the local $S$-matrix that is associated to a given Lagrangian. In the latter approach the Haag-Kastler axioms are not assumed but they are shown to follow from properties of the local $S$-matrix. This is an essential difference that we will not elaborate on, since here we are simply borrowing properties of the $S$-matrix to describe operations and we are not yet drawing a full connection to \cite{Buchholz_2020}.}
Then, thanks to microcausality 
\begin{equation}\label{eq:commuting_s}
    \left[S[f],O\right]=0, \text{ if } O\in \mathfrak{A}(\text{R})
\end{equation}
for any region R spacelike to $\supp{f}$. Equivalently, 
\begin{equation}
    S[f]^\dagger OS[f]=O.
\end{equation}
namely spacelike separated observables remain unchanged. While it may seem that this is sufficient to ensure no-signalling, this is not the case (see next subsection). As we will demonstrate, ruling out no-signalling in impossible measurements scenarios, involves zooming into the support of $f$ and making sure that the local operation can be decomposed such that future evolution cannot affect past evolution (retrocausality).


Overall, the relevant properties of the local scattering maps $S=S[f]$ that we will use below are: 
\begin{enumerate}
    \item Unitarity: $SS^\dagger = S^\dagger S= \openone$.
    \item Normalisation: $S[0]=\openone$.
    \item Locality: An $S$-matrix $S_\textsc{r}$ is local to a region R if 
    \begin{equation}\label{eq:locality}
        [S_\textsc{r},\mathfrak{A}(\text{R}')]=0
    \end{equation}
    for all regions R$'$ spacelike to R\footnote{When the local algebras are von Neumann and Haag duality holds, this condition implies that $S_\textsc{r}$ is affilated to $\mathfrak{A}(\text{R})$. Since it is assumed unitary, this is sufficient to imply that $S_\textsc{r}\in\mathfrak{A}(\text{R})$.}.
    \item Continuous additivity: given a local $S$-matrix that depends functorially on a smearing function $f$ such that $S[f]$ is local to $\supp{f}$ and any Cauchy surface $\Sigma$, we can define $f_+=f\chi_{J_+(\Sigma)}$, $f_-=f\chi_{J_-(\Sigma)}$, and have
    \begin{equation}
        S[f]=S[f_+]S[f_-]. \label{eq:continuousfactorisation}
    \end{equation}

\end{enumerate}

This property is central in our analysis of signalling and retrocausality.
It allows us to decompose the dynamical evolution in past and future evolution with respect to \textit{any} given foliation, and is the covariant analogue of $U(t_2-t_0)=U(t_2-t_1)U(t_1-t_0)$ for $t_0\prec t_1 \prec t_2$ (or the corresponding property of the path integral propagator). 
Note that by decomposing $f$ in this way, $f_\pm=f\chi_{J_{\pm}(\Sigma)}$ are not smooth due to the characteristic function. It is key that we are cutting $f$ on a Cauchy slice, i.e. a spacelike hypersurface, rather than a lightlike hypersurface such as one that coincides with A or C's lightcones, as factorising the local $S$-matrix by cutting $f$ along the lightcone would introduce divergences (which can be seen using microlocal analysis \cite{Brunetti_2000}). We expect that there are no divergences when splitting with respect to a \textit{spacelike} hypersurface for sufficiently simple interactions, and  in Appendix \ref{app:temporal_switch} provide a detailed argument for the specific case of the linear coupling that we treat below.


Finally, 
for composing distinct (causally orderable) interactions we additionally need to assume the following:
\begin{enumerate}\setcounter{enumi}{4}
    \item Causal factorisation: Given two disjoint regions A, B with B in the out region of A ($\text{A}_\text{out}=M\backslash J^-(\text{A})$, see figure \ref{fig:out_region}), the evolution is given by an $S$-matrix $S_{\textsc{a+b}}$ which factorises 
    \begin{equation} \label{disjoint operational}
        S_{\textsc{a+b}}=S_\textsc{b} S_\textsc{a}
    \end{equation}
where $S_\textsc{a}\; (S_\textsc{b})$ is the $S$-matrix where just region A(B) has non-zero coupling, and is local to A(B). 

\end{enumerate}

Note that if the supports are fully spacelike separated then it also holds that $S_{\textsc{a}+\textsc{b}}=S_\textsc{b} S_\textsc{a}=  S_\textsc{a} S_\textsc{b}$ and as a result
\begin{equation}
    [S_{\textsc{a}}, S_{\textsc{b}}]=0
\end{equation}
for A, B spacelike. \footnote{This is how microcausality is shown to hold for the local algebras generated by the local $S$-matrix \cite{rejzner2019localitycausalityperturbativeaqft, Buchholz_2020}, as a consequence of causal factorisation.}

\subsubsection{Eliminating the `impossible'}

To appreciate the operational consequences of the properties listed above, we need to define a valid notion of operation based on the local $S$-matrix. The first step is to enlarge the Hilbert space by including the Hilbert space of probe systems  that are coupled to the field through local $S$-matrices (one probe per interaction region). Consider $n$ probes and $n$ regions $\text{R}_i$, each of which are disjoint but not necessary all mutually spacelike, given by quantum systems defined on $\mathcal{H}_\text{field}\bigotimes_{i=1}^n\mathcal{H}_i$. Each of these probes can be associated to an interaction term
\begin{equation}
    \mathcal{L}_i\in \mathfrak{A}(\text{R}_i)\otimes \mathfrak{A}(\mathcal{H}_i)\bigotimes_{j\neq i} \openone_j
\end{equation}
which is a coupling between the field and one probe, localised to some region $\text{R}_i$. Approaches to measurement theory in QFT in which the measurement process is modelled as a local scattering process and operations are induced by probes, include the detector-based measurement theory \cite{Polo_Gomez_2022} and the Fewster-Verch (FV) framework in algebraic QFT \cite{fewster2020quantum}. 

In the detector-based approach, one works with (a general class) of Hamiltonian densities that model the detector-field coupling e.g. mimicking the light-matter interaction, and the local $S$-matrix can be defined as the time-ordered exponential of these interaction Hamiltonians (typically the predictions of the models rely on perturbation theory, using the Dyson expansion of the $S$-matrix). In the algebraic approach, the formulation of the FV framework relies on the structure of abstract $^*$-algebras. In this setting, the interaction between system and probe field is formulated as a morphism $\Theta$ of the uncoupled theory, and it is not necessarily implemented through the adjoint action of a unitary scattering map $S$, namely as $\Theta (\cdot)= S^{\dagger} (\cdot) S$. Note that the approach we are taking here is more abstract than the first (since it does not rely on particular interaction Hamiltonians and the Dyson expansion), but more concrete than the latter, since we assume the existence of these local scattering maps that implement the scatteting morphism in a Hilbert space.

Partial or exact resolutions to the impossible measurements scenario have been offered respectively in the detector-based approach \cite{Benincasa_2014, deramon2021relativistic} and in the FV-framework \cite{ PhysRevD.103.025017}. In either case, an important ingredient of the resolution was an analogue causal factorisation property Eq \eqref{disjoint operational}. An abstract version of causal factorisation is assumed to hold in the FV framework \footnote{The abstract version of the causal factorisation condition is that $\Theta_{\textsc{a+b}}= \Theta_{\textsc{b}} \circ \Theta_{\textsc{a}}$, which has to be proven for concrete models in this framework.}, while in \cite{deramon2021relativistic} it was argued that that this  property holds for a general class of Unruh-DeWitt-type detector models coupled locally to a coupled field over disjoint, causally orderable regions. 
Finally, note that all the measurement schemes considered in QFT correspond to so-called \textit{pre-measurement processes}, involving only the unitary interaction between the field and the probe. Strictly speaking, the measurement is completed only once an outcome has been obtained, which would require, in turn, a measurement of the probe and so on, leading to an infinite regress, see \cite{mandrysch2024quantumfieldmeasurementsfewsterverch}. In this work we ignore this final measurement stage and, following common jargon, refer to the pre-measurement stage itself as the “measurement process”.

 We should emphasise that in both approaches causal factorisation is assumed (or proven on a case-by-case basis) but, crucially, it is not sufficient for eliminating impossible measurements. More assumptions about the probe-field interactions in the coupling regions have to be made, i.e. the locality of the detector's local current in the detector model approach \cite{deramon2021relativistic} \footnote{In \cite{deramon2021relativistic} the authors show that (seemingly local) Hamiltonian densities of the form $h(\mathsf x)= \Lambda(\mathsf x)J(\mathsf x)\otimes \phi(\mathsf x)$, where $J(\mathsf x)$ is the current associated to a probe, can give rise to impossible measurements if the probe's dynamical current is not microcausal (i.e. if $[J(\mathsf x),J(\mathsf y)]\neq 0$ for $\mathsf x, \mathsf y$ spacelike separated) by using the Dyson expansion of $S=\mathcal{T}\text{exp}\,i\int h(\mathsf x) \d \mathsf x$. }, or the time-slice property of the coupled theory in the FV framework \footnote{The exact relation between continuous additivity and other dynamical axioms e.g. the time-slice property in the formal setting of (p)AQFT goes way beyond our purposes (for related results see e.g. \cite{ruep_thesis_2022}).} \cite{PhysRevD.103.025017}.
Our claim is that continuous additivity (Eq \eqref{eq:continuousfactorisation}) is the key additional assumption that gurantess that the probe-field coupling will not enable superluminal signalling. Specifically, we will see that the ability to appropriately factorise the interaction region into causally ordered subregions is \textit{sufficient} for ruling out impossible measurements. In the next section, we will see that it is \textit{not necessary}, since a weaker factorisation property can be used.

\begin{figure}
    \centering
    \includegraphics[width=1\linewidth]{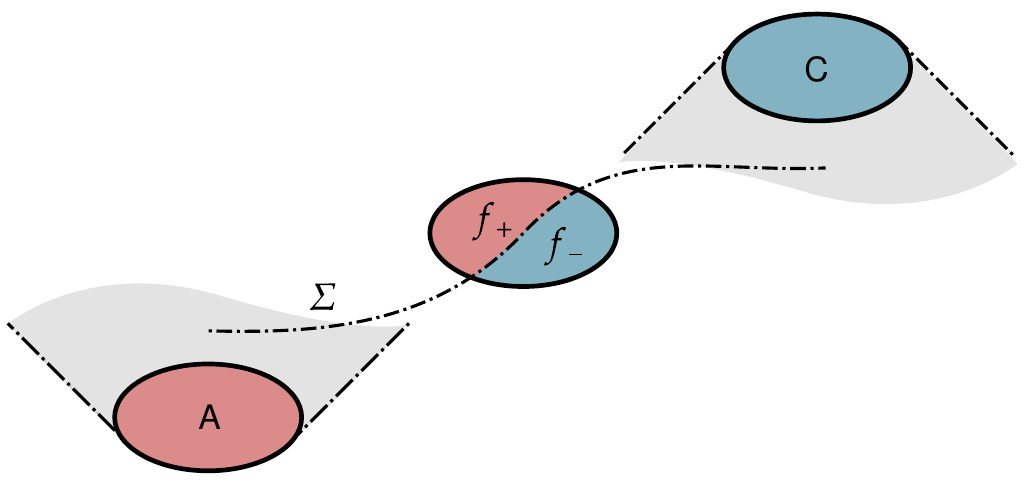}
    \caption{A Sorkin scenario where the dynamics of the interior region can be split as given in Eq \eqref{eq:continuousfactorisation}. Note that $\Sigma$ is chosen such that $\supp{f_-}$ is spacelike to A and $\supp{f_+}$ is spacelike to C.}
    \label{fig:sorkin_cont}
\end{figure}

To see that Eq \eqref{eq:continuousfactorisation} is sufficient, let us consider a Sorkin-type scenario, where we assume causal factorisation for the interactions over the coupling regions A and B, and continuous additivity for Bob's interaction over region B, see figure \ref{fig:sorkin_cont}. The total $S$-matrix is given by 
\begin{equation}
    S_{\textsc{a+b}}=S_\textsc{b}[f]S_\textsc{a}=S_\textsc{b}[f_+]S_\textsc{b}[f_-]S_\textsc{a}.
\end{equation}
Given an initial total state $\rho$ of the field and the two probes, the expectation value of a field observable $C$ in region C after both interactions is (omitting the identity operators for convenience)
\begin{align}\label{eq:cont_fact_impossible}
    \text{Tr}(S_{\textsc{a+b}}^\dagger  & CS_{\textsc{a+b}}\rho  )\nonumber \\=&\text{Tr}(S_\textsc{a}^\dagger S_\textsc{b}[f_-]^\dagger S_\textsc{b}[f_+]^\dagger CS_\textsc{b}[f_+]S_\textsc{b}[f_-]S_\textsc{a}\rho)\nonumber\\
    =&\text{Tr}(S_\textsc{a}^\dagger S_\textsc{b}[f_-]^\dagger CS_\textsc{b}[f_-]S_\textsc{a}\rho)\nonumber\\
    =&\text{Tr}(S_\textsc{b}[f_-]^\dagger CS_\textsc{b}[f_-]\rho)
\end{align}
where we have used the locality properties—i.e. $S[f]$ is localised to $\supp{f}$ and $S_{\textsc{r}}$ is localised to region R for $\text{R}=\text{A}, \text{B}$— to commute $S[f_+]$ with $C$ and $S_\textsc{a}$ with $S[f_-]$ and $C$ (see figure \ref{fig:sorkin_cont}) to then use that $S_\textsc{b}[f_+]^{\dagger}S_\textsc{b}[f_+]=\openone_{\textsc{b},  \text{field}}$ and $S_\textsc{a}^{\dagger}S_\textsc{a}=\openone_{\textsc{a}, \text{field}}$. Since there is no trace of Alice's interaction with the field ($S_{\textsc{a}}$) in the final expression, there can be no superluminal signalling from region A to region C. Note that the trace operation (Tr) is taken over $\mathcal{H}_{\text{field}}\otimes \mathcal{H}_{\textsc{a}} \otimes \mathcal{H}_{\textsc{b}}$. If we assume that Alice is initially uncorrelated with the field (and Bob) $\rho=\rho_{\textsc{a}}\otimes \rho_{\textsc{b}, \text{field}}$ and using that $\text{Tr}_{\textsc{a}}\rho_{\textsc{a}}=1$, the initial state of Alice $\rho_{\textsc{a}}$ will also drop out from the expectation value in Eq \eqref{eq:cont_fact_impossible}. 

Importantly, also note that only the part of Bob's coupling that is in contact with Charlie's causal past, i.e. $S[f_-]$ survives in the final expression, while $S[f_+]$ which is fully spacelike separated from Charlie has dropped out. Yet, for any fixed splitting of region B, the support of $f_-$ contains points that are spacelike separated from C, as the splitting is along a spacelike Cauchy surface. In principle, changes to the $S$-matrix at these points should not affect expectation values at C. This observation will become important when we introduce the notion of effective Kraus operator in section \ref{subsec:causality_for_example}.

Finally, note that introducing the probes above was only the first step towards defining a notion of operation in the next sections. We will do this by suitably taking partial traces over the probes' degrees of freedom, to induce an operation on the Hilbert space of the field, while so far we are only describing the full field-probes dynamics and calculating expectation values of field observables.

\subsection{Hierarchy of factorisation conditions and the Hammerstein property}\label{subsec:causal_fact}

So far we have introduced two notions of factorisations of $S$-matrices, causal factorisation and continuous additivity, that respectively correspond to composing disjoint interactions or decomposing a single interaction. Overall, there is a hierarchy of factorisation conditions, that we will briefly outline here for completeness, and generalise our proof of no impossible operations to a weaker condition. 

Firstly, we note that Eq \eqref{eq:continuousfactorisation} is quite a stringent condition on the local $S$-matrix. This condition might fail for certain singular interactions in certain spacetime dimensions\footnote{A related axiom postulated by Schwigner is that there exist unitaries such that $U_{\Sigma, \Sigma''}=U_{\Sigma,\Sigma'}U_{\Sigma'\Sigma}$, where $\Sigma'$ a spacelike hypersurface between $\Sigma$ and $\Sigma''$ hypersurfaces. This is known to hold for polynomially interacting scalar fields in two dimensions.}. For this reason, here we will also consider weaker factorisation conditions that are more widely assumed to hold for local interactions.

One such weaker condition, called the \textit{Hammerstein property}, is typically assumed in causal perturbation theory literature \cite{rejzner2019localitycausalityperturbativeaqft,Brunetti_2000}. The Hammerstein property is that 
\begin{equation}\label{Hammerstein}S[f_1+f_2+f_3]=S[f_1+f_2]S[f_2]^{-1}S[f_2+f_3] 
\end{equation}
where $f_1,f_3$ are separated by a hypersurface and $f_2$ is unrestricted (and it can potentially overlap with $f_{1,3}$). The test functions $f_1,f_2,f_3$ are assumed to be smooth and compactly supported. This implies causal factorisation for $f_2$ being the zero function  
\begin{equation}\label{eq:weak}
    S[f_1+f_3]=S[f_1]S[f_3]
\end{equation}
where the supports of $f_1,f_3$ are causally orderable, i.e. can be separated by a Cauchy surface. Note that continuous additivity 
\begin{equation} \label{additive}
    S[f]=S[f_+]S[f_-]
\end{equation}
where $f=f_++f_-$ (and $f_\pm$ need not be smooth at the hypersurface $\Sigma$) in turn implies the weaker condition Eq \eqref{Hammerstein}. To see this, we split $f$ into smooth test functions $f'_+,f'_0,f'_-$ configured as required for Eq \eqref{Hammerstein} and such that $f=f'_++f'_0+f'_-$. Applying condition Eq \eqref{additive} twice,
\begin{align}
    S[f]=&S[f'_++f'_0-f'_0+f'_0+f'_-]\nonumber\\=&S[f'_++f'_0]S[-f'_0] S[f'_0+f'_-]\nonumber\\=& S[f'_++f'_0]S[f'_0]^{\dagger} S[f'_0+f'_-]
\end{align}
where we have used that $S[0]=S[f]S[f]^\dagger\Rightarrow S[f]^\dagger =S[-f]$.

Importantly, the Hammerstein property is also sufficient to rule out impossible measurements, as we explicitly show in detail in Appendix \ref{app:hammerstein} and sketch here. 
\begin{figure}
    \centering
    \includegraphics[width=1\linewidth]{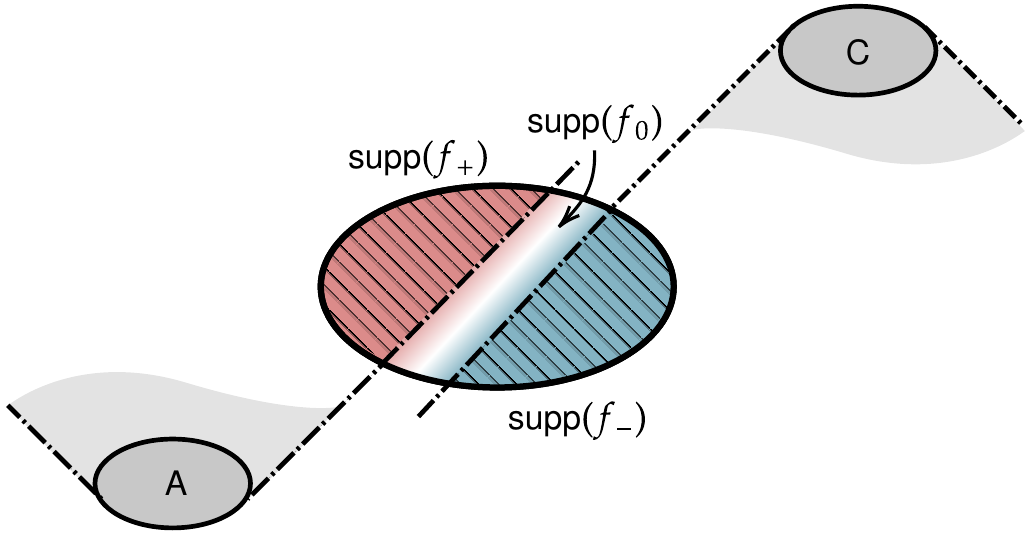}
    \caption{The spacetime factorisation of $f$ into $f_+$ (red) $f_0$ (white) and $f_-$ (blue), where we note the disjoint and time ordered supports of $f_\pm$, and the support overlaps of $f_0$ with $f_\pm$. Key to the proof is that the support of $f_0$ is spacelike separated from both A and C. }
    \label{fig:hammerstein_sketch}
\end{figure}
Looking at figure \ref{fig:hammerstein_sketch}, we can see a similar scenario to the proof for the stronger factorisation condition. As there, the supports of $f_\pm$ are spacelike to regions $\text{C},\text{A}$ respectively, and now $f_0$ is spacelike to A and C. This implies that $S[f_++f_0]S[f_0]^\dagger$ commutes with $C$ and $S[f_-+f_0]^\dagger CS[f_-+f_0]$ is localised spacelike to A. Following a very similar derivation to Eq \eqref{eq:cont_fact_impossible} demonstrates that there is no signalling in a Sorkin scenario where $\Phi_\textsc{b}$ is induced by an $S$-matrix satisfying the Hammerstein property \footnote{The resolution of the Sorkin problem in the abstract algebraic setting of the FV framework heaviy relies on the time-slice property of the coupled theory \cite{PhysRevD.103.025017, papageorgiou2024eliminating}. At the level of the local $S$-matrix it has been shown that the time-slice property can be derived only from \eqref{Hammerstein} and normalisation $S(0)=\openone$ \cite{chilian2009time}, which explains why we were able to demonstrate the link between \eqref{Hammerstein} and no impossible operations in Appendix \ref{app:hammerstein}.}. For the full details, see Appendix \ref{app:hammerstein}.

To summarise, the weakest factorisation condition, causal factorisation, is necessary but not sufficient for resolving the Sorkin problem, while the strongest, continuous additivity, together with causal factorisation completely rules out Sorkin's impossible operations. So what is the role of the middle condition Eq \eqref{Hammerstein}? It is definitely reassuring that Eq \eqref{Hammerstein} rules out impossible measurements when continuous additivity fails (Eq \eqref{eq:continuousfactorisation}). Nevertheless,
since the concrete model for scalar field measurements that we consider in the next section satisfies the stronger, and simpler, condition Eq \eqref{eq:continuousfactorisation}, we will utilise that instead for translating it into properties of induced Kraus operators. 

\section{Operations induced by probes} \label{sec:Operations induced by probes}

In quantum mechanics, every unital, i.e. maps the identity to itself, CP map (UCP map) -- that is, every non-selective operation on a quantum system -- can be understood as arising from a unitary interaction between the system and an ancillary system, or probe, followed by discarding (tracing out) the probe. This is a consequence of the Stinespring dilation theorem, which ensures that any UCP transformation of the system can be ``purified'' by embedding it into a larger closed system that evolves unitarily. In the same vein, we will consider in this section the realisation of non-selective operations in QFT through interactions between the system (i.e., the field) and a probe. Then, a way to argue that unitary ``kicks'' which violate no-signalling are not physically realisable is to show that they cannot be implemented by any valid local interaction between the system and a probe (see also \cite{muchverch}).

 Below, we define local Kraus operators as induced by a suitable local  coupling $S[f]$ between the field and the probe, and we provide an example of a concrete model for field measurements. We derive no-signalling conditions for the associated Kraus operators.

\subsection{Induced Kraus operators in QFT}\label{subsec:kraus_operators_qft}


The attempts of modelling local operations in QFT required a departure from the asymptotic scattering paradigm. In a series of papers, Hellwig and Kraus \cite{hellwigkraus1969,hellwigkraus1970b} model relativistic quantum measurement as a local scattering process. Their motivation for introducing the so-called Kraus operators, as induced through the dynamical coupling to some probe, was to explicate the operational interpretation of the local algebras that is implicit in the algebraic approach to QFT \cite{haag_kastler_1964}. The measurement scheme of Hellwig and Kraus can be viewed as predecessor of the recently proposed measurement frameworks in QFT \cite{Polo_Gomez_2022,fewster2020quantum} that involve some (abstract or concrete) representation of dynamics between system and probe and model measurement as a local scattering process (for historical context see \cite{Fraser2023}). 
 
Consider the combined Hilbert space $\mathcal{H}_{\text{field}} \otimes \mathcal{H}_{\text{probe}}$ of a system field with Hilbert space $\mathcal{H}_{\text{field}}$ and some probe with Hilbert space $\mathcal{H}_{\text{probe}}$. Using interaction picture the final result of the probe-system interaction can be described through a unitary operator $S$ on the product space $\mathcal{H}_{\text{field}}\otimes\mathcal{H}_{\text{probe}}$. We will assume that the coupling $S$ is supported over a bounded spacetime region, i.e., $S=S[f]$ where $f=f(\mathsf x)$ is a spacetime smearing function that is compactly supported. 

We will only use the the dual update map, acting on the operators rather than the states. This is most natural as the QFT operators we are working with are in the Heisenberg picture. The update rule is now given by 
\begin{equation}
    O\mapsto \int \d q {\Pi^{\psi}_q[f]}^\dagger O \Pi^{\psi}_q[f]=:\Phi[f](O)  \label{HK}
\end{equation}
where we have included the functional dependence on the spacetime smearing function $f$, and the Kraus operators for the field associated to $S[f]$ are given by
\begin{equation}
    \Pi^{\psi}_q[f] := \braket{q| S[f]|\psi}\label{kraus},
\end{equation}
where $\psi,q$ denote the initial and final states of the probe only, so $\Pi^\psi_q[f]$ is an operator on $\mathcal{H}_{\text{field}}$. This can be derived by taking the partial trace over the probe's Hilbert space (for details see e.g. \cite{Polo_Gomez_2022}). It is easy to check that this defines a valid family of Kraus operators that satisfy
\begin{equation}
  \int \d q   \Pi^{\psi}_q[f]^\dagger \Pi^{\psi}_q[f] =\openone_{\text{field}}
\end{equation}
thanks to the unitarity of the local $S$-matrix and $\psi$ being a normalised state, using the resolution of identity $\int \d q |q\rangle \langle q|=\openone_{\text{probe}}$ in the probe's Hilbert space.

Kraus operators in scattering-based frameworks for local measurement in QFT have been derived previously in \cite{SmithAlexander2017, Polo_Gomez_2022, PhysRevD.109.065024} for concrete Unruh-DeWitt-type models, and more recently in \cite{mandrysch2024quantumfieldmeasurementsfewsterverch} for a concrete representation of the FV framework in a Hilbert space. 
Nevertheless, the operational consequences of the factorisation conditions at the level of Kraus operators have not been investigated. Here we investigate how these properties are captured by valid local scattering matrices and their (de)composition properties. Further, for certain concrete model for measurement of field amplitudes, we link these properties to a no-signalling property for the resulting Kraus operators.

\subsection{A concrete model for local field measurements}\label{sec:example}

We consider a simple model for field measurements, previously explored in \cite{PhysRevD.109.065024}, where a scalar quantum field $\phi$ in Minkowski spacetime interacts with a free pointer via the operator $P$, satisfying $[Q, P] = i$. The interaction is defined by the following coupling:
\begin{equation} \label{coupling}
\mathcal{L}_I[f] = f(\mathsf{x}) \phi(\mathsf{x}) \otimes P. 
\end{equation}
 Here, $f(\mathsf{x})$ is a spacetime-dependent smearing function. The pointer's free Hamiltonian can be chosen to be non-trivial as long as it commutes with the interaction term, without introducing significant computational challenges, so it is theoretically possible to select the Hamiltonian of a relativistic scalar particle. However, to avoid conceptual complications associated with interpreting the pointer as a relativistic particle—such as issues of causality and localization in relativistic quantum mechanics (e.g., Hegerfeldt's \cite{hegerfeldt2001particlelocalizationnotioneinstein} and Malament's \cite{malament1996defense} theorems)—we opt to set the pointer's Hamiltonian to $H_\text{pointer} = 0$. This approach treats the pointer as an abstract degree of freedom, not inherently tied to spacetime except through its localized coupling to the field. To reflect this distinction, we will use $\mathsf{x}$ to denote spacetime points and $q$ to represent the internal values of the pointer. The interaction in Eq \eqref{coupling} leads to the following scattering matrix
\begin{equation} \label{scatteringmap}
    S[f]=\mathcal{T}e^{-i\phi(f)\otimes P}
\end{equation}
which we can solve exactly via the Magnus expansion \cite{PhysRevD.109.065024}. 

For this model the $S$-matrix takes the form
\begin{equation}\label{quantum_controlled}
    S[f]=\mathcal{T}e^{-i\phi(f)\otimes P}=\int dp \mathcal{T}e^{-ip\phi(f)}\otimes \ket{p}\bra{p},
\end{equation}
where by $p$ we denote the spectrum of the continuous pointer $P$. The Magnus expansion allows us to rewrite time ordered exponentials $\mathcal{T}\text{exp}(\cdot)= \text{exp} \sum_n \Xi_n(\cdot)$, where $\Xi_n$ are given in terms of $n$-th order nested commutators. Since the pointer evolves trivially, the Lagrangian term $\mathcal{L}(x)=\phi(x)\otimes P$ implies 
\begin{equation}
    [\mathcal{L}(x),\mathcal{L}(x')]=c\openone,
\end{equation} 
and so only the terms that include at most one commutator survive,
\begin{equation}
    \Xi_1=-ip\phi(f),\; \Xi_2=-\frac{ip^2}{2}\Delta_{\textsc{r}}(f,f),\; \Xi_n=0, \; n>2,
\end{equation}
where $\Delta_{\textsc{r}}$ is the retarded propagator. Finally, we have
\begin{equation}
    S[f]=\int \d p e^{-p^2\frac{i}{2}\Delta_{\textsc{r}}(f,f)} \left(e^{-ip\phi(f)}\otimes \ket{p}\bra{p}\right). \label{Scontroled}
\end{equation}
and
\begin{equation}
   \Delta_{\textsc{r}}(f,f)= \int \d \mathsf x \d \mathsf y f(\mathsf x) \Delta_{\textsc{r}}(\mathsf x,\mathsf y) f(\mathsf y).
\end{equation}

After picking a probe state $\psi(q)$ on the basis of the pointer variable $Q$, using Eq \eqref{kraus} we can compute the Kraus operators for the field \cite{PhysRevD.109.065024}
\begin{align} \label{Kraus}
    \Pi_q^\psi[f]=&\frac{1}{\sqrt{2\pi}}\int \d p\, \tilde{\psi}(p)\nonumber\\&\times\text{exp}\left(-\frac{i}{2}p^2\Delta_{\textsc{r}}(f,f)+ip(q-\phi(f))\right), 
\end{align}
where $\tilde{\psi}(p)=\langle p| \psi\rangle$ is the Fourier transform of $\psi(q)$. Note that the expression \eqref{Kraus} has the form of a Fourier transform with an additional phase that depends on $\Delta_{\textsc{r}}(f,f)$, which arises due to the time-ordering operator in Eq \eqref{scatteringmap}, using the Magnus expansion \footnote{The retarded propagator is due to the time-ordering operator. The deeper reason why we end up only with retarded (and not the advanced) propagation is the implicit time-asymmetry introduced by assuming initial (and not final) separable state.}. It is precisely the causal retarded propagation within the support of $f$ that this model captures at the level of the Kraus operators, which will play an essential role in analysing the resolution of field measurements, and the no-signalling properties of the Kraus operators in the next section. 
 
As an example, by picking $\psi(q)=G_\sigma^{1/2}(q)$ to be the square root of the Gaussian centred at $0$  with width $\sigma$, we have the following POVMs
\begin{align}
  {\Pi^{\sigma}_q[f]}^\dagger \Pi_q^{\sigma}[f]&= \frac{1}{\sqrt{2\pi} \tilde\sigma} \text{exp} \left[-\frac{\left(q-\phi(f)\right)^2}{2\tilde\sigma^2} \right]\nonumber\\&=G_{\tilde\sigma}(q-\phi(f))
\end{align}
where 
\begin{equation}
    \tilde\sigma=\sqrt{\sigma^2+\frac{\Delta_{\textsc{r}}(f,f)^2}{4\sigma^2}}.
\end{equation}

Using the inequality between the arithmetic and geometric means, we see that 
\begin{equation} \label{bound}
    \sigma^2+\frac{\Delta_{\textsc{r}}(f,f)^2}{4\sigma^2}\geq 2\sqrt{\frac{\Delta_{\textsc{r}}(f,f)^2}{4}}=|\Delta_{\textsc{r}}(f,f)|
\end{equation}
and so the minimal value of $\tilde\sigma$ is $|\Delta_{\textsc{r}}(f,f)|$ and occurs at $\sigma^2=|\Delta_{\textsc{r}}(f,f)|/2$. Note that this POVM governs the statistics of the conjugate pointer variable $Q$ after the coupling is `off' (that is, in the out region of $\supp{f}$).

The uncertainty $\tilde\sigma$ represents how far the POVM is away from being a projection valued measure (PVM), or in quantum measurement language, how far from sharp the measurement is. It is not hard to see that if we could take $\tilde{\sigma}\rightarrow 0$, then the POVM would become
\begin{align}
    &\Pi^{\sigma}_q[f]^\dagger \Pi_q^{\sigma}[f]\rightarrow \delta(q-\phi(f))\nonumber\\=&\int \d \lambda \delta(q-\lambda)dP(\lambda)=P(q)
\end{align}
where $P(q)$ is the PVM associated to the possible values of $\phi(f)$ by the spectral theorem.

\subsection{Can $\phi(f)$ be measured sharply?}

In \cite{ozawa1993canonical}, Ozawa studies the process of taking such a limit, and shows that a measurement is only repeatable in that limit, in the sense that the outcomes of repeated measurements agree. Operationally, they consider allowing the interaction to continue for a longer time, say a factor of $T$, and rescaling the pointer units by $1/T$. It is shown in the case considered in that paper the standard deviation is scaled by $1/T$. This is mathematically the same as scaling the probe's state's argument $q$ by $T$, thus scaling $\sigma$ by $1/T$. We cannot do quite the same here, as we now discuss.

Suppose we take $\sigma$, the probe's standard deviation in $q$, to zero, then the first term in Eq \eqref{bound} will indeed vanish, however, the second term will explode, meaning the uncertainty diverges, $\tilde\sigma\rightarrow \infty$. The appearance of the propagator term in Eq \eqref{bound} is a consequence of solving the Dyson series for the $S$-matrix exactly, rather than assuming the interaction Hamiltonian dominates the field's free Hamiltonian. We can also see that the square root of the propagator bounds the uncertainty from below, and so we must simultaneously reduce $|\Delta_{\textsc{r}}(f,f)|$ and take the probe state to be sharper. If done such that the inequality in Eq \eqref{bound} is saturated at each step, then the POVM will become sharper and sharper. However, this requires $|\Delta_{\textsc{r}}(f,f)|$ be taken smaller. Next, we interpret this bound further.

\begin{figure}
    \centering
    \includegraphics[width=1\linewidth]{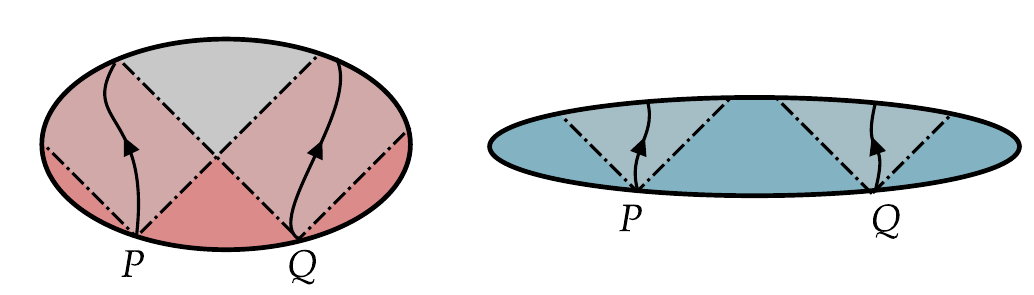}
    \caption{(Left) a region that has a large time extension will contain overlapping futures of points that are at large spacial separations e.g. $J^+(P)\cap J^+(Q)$, indicated by the darkest grey region. Conversely, if the degrees of freedom of the theory travel much slower than $c$, these overlap regions will not be important, indicated by the two black timelike curves. (Right) A region with small time extension will only have points that are close together have overlapping futures contained in itself: $J^+(P)\cap J^+(Q)=\emptyset$.}
    \label{fig:causal_influence}
\end{figure}

To gain some intuition on how accurately the field values can be probed locally, we can appeal to the results of \cite{PhysRevD.109.065024}, where it is shown that for $f$ given by a Gaussian with spacial width $\sigma_S$ and time width $\sigma_T$, 
\begin{equation}
    \Delta_{\textsc{r}}(f,f)\sim\begin{cases}
        \frac{1}{\sigma_T\sigma_S^3} & \text{ when } m\sigma_T\gg 1 \\ 
        \frac{\sigma_T}{\sigma_S(\sigma^2_S+\sigma_T^2)} & \text{ when } m=0.
    \end{cases}
\end{equation}
The intuition is that the effect of interacting with the field will disturb it, and those disturbances will propagate following the dispersion relations. When $m\gg \frac{1}{\sigma_T}$, those disturbances will travel non-relativistically, and so will not probe the spacial extent of the region $f$, while when $m\sim 0$, they will travel close to or at the speed of light, see figure \ref{fig:causal_influence}. 

The retarded propagator encapsulates how much a perturbation at $\mathsf x$ affects the field at $\mathsf y$. For all masses, it vanishes if the points are spacelike separated, and if the mass is zero then it diverges when there is a lightlike curve from $\mathsf x$ to $\mathsf y$, meaning that the perturbation caused by the measurement interaction at $\mathsf x$ will propagate along the lightcone and affect the measurement at $\mathsf y$. If the region is spacially wide but temporally narrow, i.e. the region is not well causally connected, then not much of the support of $f$ is contained in the future lightcone of any point $\mathsf x$, and so there is minimal disturbance. Importantly, if the smearing is purely spacial, that is in the case of instantaneous interactions, then $\Delta_{\textsc{r}}(f,f)=0$. This allows us to draw a connection to the operations that appear in \cite{oeckl2024spectraldecompositionfieldoperators} (see next section). 

Conversely, if the region is spacially narrow and temporally long, then this is flipped on its head, and the disturbance is large, see figure \ref{fig:causal_influence}. Finally, if the mass of the particle is non-zero, then the retarded propagator is no longer concentrated on the lightcone, as represented by the timelike worldlines in figure \ref{fig:causal_influence}. In this case, the causal connectivity of the region is less important, and the disturbance instead depends on the overall spacetime volume of the region. In the $m\gg \frac{1}{\sigma_T}$ case, increasing either size of $\sigma_T, \sigma_S$ makes the measurement sharper, while for $m\sim 0$, when $\sigma_S^2+\sigma_T^2$ is held constant, the region must be taken to be much wider spacially than temporally in order to get a sharp measurement. 

Overall, we see that explicitly modeling the measurement interaction leads to a quantum channel with a sharpness bound that stems from the $\Delta_{\textsc{r}}(f,f)$ phase in the $S$-matrix. This phase arises from solving the dynamics of the field and pointer, rather than from imposing causality constraints. While the $\Delta_{\textsc{r}}(f,f)$ phase is not necessary to prevent signaling in the field channel (see Eq \eqref{eq:second_sorkin} where the unitaries that define the channel are not time-ordered), it is critical for proving continuous additivity and ensuring the causality of the dilated channel (see Appendix \ref{appendix:additivity} and \ref{causal convolution}). Finally, we comment on the connection to previous related work on families of no-signalling Kraus operators and POVMs for field measurements. Our analysis provides a concrete dilation of the $L^2$-maps that defines families of Kraus operators in \cite{albertini2023ideal}, defined as 
\begin{equation}
    K^f_\kappa(q)=\kappa(q-\phi(f)).
\end{equation}
By picking the internal state of the probe to be
\begin{equation}
    \psi(q):=\Psi_f(\kappa)[q]:=\mathcal{F}^{-1}\left(\mathcal{F}(\kappa)[p]e^{\frac{ip^2}{2}\Delta_{\textsc{r}}(f,f)}\right) [q]
\end{equation}
where $\mathcal{F},\mathcal{F}^{-1}$ are the Fourier and inverse Fourier respectively, we can see that 
\begin{equation}
    K^f_\kappa(q)=\Pi^{\Psi_f(\kappa)}_q[f].
\end{equation}
This means that the `shape' of the Kraus operators is specified by the initial state preparation $\psi$ of the probe. At last but not least, in \cite{mandrysch2024quantumfieldmeasurementsfewsterverch} the authors analyse a concrete model of the FV-framework that involves two equal mass real scalar fields linearly coupled, where one of the two fields (the probe field) plays the role of the `pointer'. By picking a squeezed quasi-local state for the probe field, they show that a Gaussian measurement of the probe leads to an instrument that in the weak coupling limit can be decomposed into a Gaussian measurement and a dephasing channel with certain strength. A full comparison between these results and ours would require letting the masses of both fields, the probe and the system field, to vary independently. This is because our pointer is a single abstract degree of freedom, and has trivial internal dynamics, which would appear in a well controlled large mass limit of the probe represented as a quantum field, likely requiring us to reintroduce a probe Hamiltonian. We leave further exploration for future works. 
\section{Causality conditions on the Kraus operators}\label{subsec:causality_for_example}

In this section we will examine the relevance of the continuous additivity Eq \eqref{eq:continuousfactorisation}  for the concrete model of field measurements that we introduced in the previous section. We will demonstrate the model's consistency with the framework laid out in section \ref{subsec:rel_dynamics}, and derive the no-signalling and compositional properties of its Kraus operators, generalising those found in \cite{oeckl2024spectraldecompositionfieldoperators}. These no-signalling identities ensure that this model does not induce impossible operations on the field.


First, we show that for the model introduced in the previous section continuous additivity holds, i.e., for any Cauchy slice through $f$ (see figure \ref{fig:sorkin_cont})
\begin{equation}\label{eq:example_additivity}
    S[f]=S[f_+]S[f_-],
\end{equation}
where $\supp f_+ \notbackslash{\preceq} \supp f_-$. We derive this property in Appendix \ref{causal convolution} using the Magnus expansion 
\begin{align}
    S[f]&=\mathcal{T}e^{-ip\phi(f)}\nonumber \\
    &= e^{-\frac{ip^2}{2}\Delta_{\textsc{r}}(f,f)} e^{-ip\phi(f)}
\end{align}
and the simple fact that
\begin{equation}
  \Delta_{\textsc{r}}(f_+\!\,,f_-)=0,  
\end{equation}
i.e, there is no propagation from the future to the past. Note that $f_\pm$ are in general not smooth, i.e. not elements of $C^\infty_0(\mathcal M)$, and so $S[f_\pm]$ and $\Delta_\textsc{R}(f_\pm, f_\mp)$ are not immediately well defined.  In Appendix \ref{app:temporal_switch} we show that suitably introducing a discontinuous test function when writing  $S[f_\pm]$ is not problematic.

Crucially, this has the following consequence for the Kraus operators, which we prove in Appendix \ref{causal convolution}- that they obey 
\begin{equation}\label{eq:example_kraus_factorisation}
    \Pi_q^{\psi}[f]=\int \d s\, \Pi_{q-s}^{\psi_+}[f_+]\Pi_s^{\psi_-}[f_-],
\end{equation}
where $f=f_+\!+f_-$ and $\psi= \psi_+\star\psi_-$, where $\star$ is convolution. This is a way of decomposing a single time-extended measurement process, for different choices of spacelike hypersurfaces going through the region. Note that $\psi^\pm$ are not necessarily square normalised and do not represent valid states for the probe, so $\Pi_q^{\psi_{+}}[f_{+}], \Pi_q^{\psi_{-}}[f_{-}]$ do not (yet) define valid (normalised) Kraus families (see Eq \eqref{effectiveoperation} in the next subsection). 

Identity Eq \eqref{eq:example_kraus_factorisation} will be key in showing that the such an operation cannot enable signalling between spacelike separated regions in a Sorkin-type  \footnote{To gain an intuition about the form of Eq \eqref{eq:example_kraus_factorisation} it is easy to check that it holds for the spectral decomposition of the sum of two commuting self-adjoint observables. For instantaneous operations considered in \cite{oeckl2024spectraldecompositionfieldoperators} the two commuting observables correspond to the scalar field commuting with itself on spacelike separated subregions on a Cauchy slice. In our case, Eq \eqref{eq:example_kraus_factorisation} is a generalisation of this fact for non-commuting and non-self adjoint Kraus operators. The key is that they only need to commute up to a phase (see derivation in appendix \ref{causal convolution}). }. In Appendix \ref{app:identity} we prove it for a simple model for $\phi(f)$ measurements, but more generally, it follows from the continuous additivity of $S[f]$ and the fact that $S[f]$ can be seen as a quantum controlled unitary (see subsection \ref{subsec:causal_properties_kraus}). 
This provides us with a concrete example of the relationship between no-signalling conditions for dilated and undilated maps. 

Identity Eq \eqref{eq:example_kraus_factorisation} was proved in \cite{oeckl2024spectraldecompositionfieldoperators} for an $\epsilon$-family of POVMs, including the spectral decomposition of $\phi(f)$. in the case of instantaneous operations defined over a Cauchy surface ($f$ being a purely spacial smearing) and it was proved to be sufficient for enforcing no-signalling condition Eq \eqref{transparency1} in this case. Note that our results reduce to those in \cite{oeckl2024spectraldecompositionfieldoperators} when we take $\psi$ to be Gaussian with width $\varepsilon$ and $f$ is supported on a spacial hypersurface so that $\Delta_{\textsc{r}}(f,f)=0$. In our case, the concrete pointer model describes time-extended operations of field amplitudes, so \eqref{eq:example_kraus_factorisation} turns out to not be sufficient to prove no-signalling. For the proof of no-signalling, we will also need the following identity\footnote{ The connection between this work and a generalised, i.e. `time extended', version of Oeckl's results in \cite{oeckl2024spectraldecompositionfieldoperators} using a covariant path-integral description \cite{oeckl2025causalmeasurementquantumfield} is work in progress (see also \cite{oeckl2025localcompositionalmeasurementsquantum}).}. 
\begin{equation} \label{identity2}
     \int \d q \Pi^\psi_{q-s}[f]^\dagger\Pi^{\psi}_{q-s'}[f]= F_{\psi} (s-s') \openone
\end{equation}
where
\begin{equation}
    F_{\psi} (s-s')=\int \d p \, |\widetilde{\psi}(p)|^2e^{ip(s-s')}, \label{prefactor}
\end{equation}
which we derive in Appendix \ref{app:identity}.

\subsection{No impossible operations}

To see this, consider a Sorkin scenario where Alice applies $\Phi_\textsc{a}$, Bob applies $\Phi_\textsc{b}[f]$---where we have been explicit about only Bob's map's functional dependence---and Charlie measures the expectation value of $C$. Picking a Cauchy hypersurface $\Sigma$ such that A is in the future and C is in the past, we split $f=f_++f_-$ (see figure \ref{fig:sorkin_cont}). We then evaluate the expectation value of $C$
\begin{widetext}
\begin{align}\label{eq:effective}        
\text{Tr}\left(\Phi_\textsc{a}\circ\Phi_\textsc{b}[f](C)\rho\right)=&\text{Tr}\left(\Phi_\textsc{a}\circ \int dx \Pi^\psi_q[f]^\dagger C\Pi^\psi_q[f]\,\rho\right)
\nonumber\\  
=&\int \d q \d s \d s'\text{Tr}\left(\Phi_\textsc{a}\circ \Pi^{\psi_-}_{s}[f_-]^\dagger \Pi^{\psi_+}_{q-s}[f_+]^\dagger C\Pi^{\psi_+}_{q-s'}[f_+]\Pi^{\psi_-}_{s'}[f_-] \rho \right)\nonumber\\ 
 =&\int \d q \d s\d s'\text{Tr}\left(C\Pi^{\psi_-}_{s'}[f_-]\rho\Pi^{\psi_-}_{s}[f_-]^\dagger\Phi_\textsc{a}\left(  \Pi^{\psi_+}_{q-s}[f_+]^\dagger \Pi^{\psi_+}_{q-s'}[f_+]\right)\right)\nonumber \\ =&\int \d s\d s'\text{Tr}\left( F_{\psi_+} (s-s')\Pi^{\psi_-}_{s}[f_-]^\dagger C\Pi^{\psi_-}_{s'}[f_-]\rho \right) 
\end{align}
\end{widetext}

In the second line above we have used the identity Eq \eqref{eq:example_kraus_factorisation} to split each Kraus operator into those localised before and after the Cauchy slice. In the third line above we have used the localisation properties to commute $C$ with $\Pi[f_+]$ and $\Phi_{\textsc{a}}$ with $\Pi[f_-]$ (see figure\ref{fig:sorkin_cont}). The fourth line follows from the fact that the argument of $\Phi_\textsc{a}$, integrated over $q$, is proportional to the identity i.e. using Eq \eqref{identity2} which is shown in Appendix \ref{app:identity}. Importantly, this leads to $\Phi_\textsc{a}$ dropping out from the expression, and as a result the details of A's preparation disappear from C's result.

We see that in the last line of Eq \eqref{eq:effective} only the Kraus operators that are supported in $f_-$ (and not $f_+$) survive. Nevertheless, this does not have the form of an operation $\Phi_{\textsc{b}}[f_-]$, because of the double integration over $s$ and $s'$. This is because the prefactor $F_{\psi}(s-s')$ defined in Eq \eqref{prefactor} is generally different than $\delta(s-s')$ for physical states $\psi$ of the pointer. Yet, we can rearrange things to show that 
\begin{align} \label{effectiveoperation}
    \int &\d s\d s' F_{\psi_+} (s-s')\Pi^{\psi_-}_{s}[f_-]^\dagger C\Pi^{\psi_-}_{s'}[f_-] \nonumber \\ 
    &= \int \d q \Pi_q^\psi[f_-]^\dagger C\Pi_q^\psi[f_-],
\end{align}
as we explicitly demonstrate in Appendix \ref{app:effective}. This motivates the notion of effective Kraus operators in the next section. 

Overall, combining Eq \eqref{eq:effective} and Eq \eqref{effectiveoperation} we have shown that   
\begin{equation} \label{almosteffective}
\text{Tr}\left(\Phi_\textsc{a}\circ\Phi_\textsc{b}[f](C)\rho\right)=  \text{Tr}\left(  \Phi_\textsc{b}[f_-](C)\rho\right).
\end{equation}
This is a no-signalling condition for B, since it is expected to hold for all maps $\Phi_{\textsc{a}}$ defined above, and all observables $C$ and states $\rho$ of the field. The properties of the local scattering map that corresponds to B ensure that only $\Phi_\textsc{b}[f_-]$ will affect the expectation value of any observable $C$ in the causal future.

\subsection{Effective Kraus operators}

\begin{figure}[t!]
    \centering
\includegraphics[width=1\linewidth]{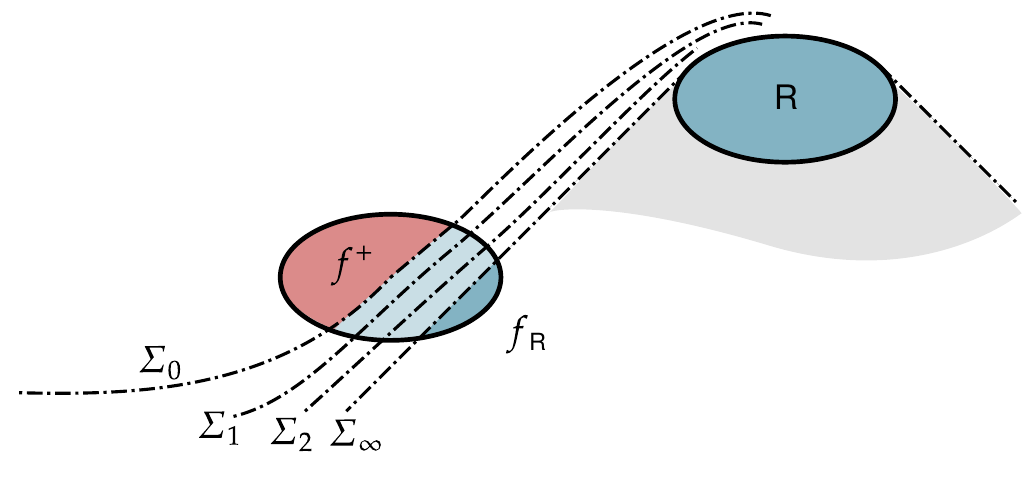}
    \caption{ The family of Cauchy surfaces $\Sigma_t$ can be seen to tend towards the intersection of the past null cone of R and $\supp{f}$.}
    \label{fig:cauchy_slice}
\end{figure}

In the previous subsection we proved Eq \eqref{almosteffective}, namely that the expectation values of $C$ do not depend on $\Phi_\textsc{a}$ and the details of B's operation over the support of $f_+$. Yet, notice that we could have used any spacelike hypersurface for splitting $f$ as $f_++f_-$. For any splitting, there are some points of $f_-$ that are spacelike to the future region R (see figure \ref{fig:cauchy_slice}) which can be excluded by using another hypersurface that is `closer' to the backward lightcone of R. This suggest that we can use covariance to get a stronger property of the Kraus operators, namely that they can be chosen to depend only on operators in the causal past of the region the algebra they act on is localised to, in a suitable limit.   This holds for general regions, so observables localised to R will be updated by \textit{effective} Kraus operators depending only on causally available data, independent of the hypersurface splitting we have used above. 

Heuristically, the limit can be defined as follows: take a continuous family of hypersurfaces $\Sigma_t$ that approach the boundary of $J_-(\text{R})$ as $t\rightarrow \infty$, in such a way that they are all in the future of $J_-(\text{R})$ at all points and $\Sigma_0$ coincides in $\supp{f}$ with the hypersurface picked for the splitting into $f_+,f_-$, see figure \ref{fig:cauchy_slice}. Then let $\chi_t$ be the characteristic function of $J_-(\Sigma_t)$, and $f_t(x)=f(x)\chi_t(x)$, and hence $f_0=f_-$, $\supp{f_t}\subset \supp{f_{t'}}$ if $t>t'$. It is clear that $f_t\rightarrow f\chi_{J_-(R)}$ as $t\rightarrow \infty$.

Finally, we use that the map $f\mapsto \Pi[f]$ is strongly continuous \cite{oeckl2024spectraldecompositionfieldoperators}\footnote{technically, this was proven only for the map between solutions and Kraus operators with $\psi$ Gaussian (see discussion of \cite{oeckl2024spectraldecompositionfieldoperators} in section \ref{sec:example}).}, to see that observables localised to R are updated by the effective `would-be' operators  
\begin{equation} \label{effective}
    \Pi_q^\psi[f_\textsc{r}], \;\; f_\textsc{r}(x)=\begin{cases}
        f & x\in J_-(\text{R})\\
        0 & x\notin J_-(\text{R}),
    \end{cases}
\end{equation}
depend only on the information causally available to R, and are independent of the exact choice of splitting. Note that these operators still correspond to the same measurement on the probe, in the sense that $\Pi_q^\psi[f], \Pi_q^\psi[f_-]$ are both conditional on the same value of the probe state $q$, and agree when acting on $\mathfrak{A}(\text{R})$. In Appendix \ref{app:temporal_switch} we sketch an argument that, at least for the simple Kraus operator form here, this limit is well-defined. For more complex Kraus operators it is not clear that such a limit can be defined. In the case of a well-defined limit, the effective update map is 
\begin{equation}
    E_f(O)=
        \int \d q \Pi_q^\psi[f_\textsc{r}]^\dagger O \Pi_q^\psi[f_\textsc{r}]
\end{equation}
where we have that 
\begin{align}
    E_f(O)=&\int \d q \Pi_q^\psi[f_\textsc{r}]^\dagger O \Pi_q^\psi[f_\textsc{r}]\nonumber\\=O&\int \d q \Pi_q^\psi[f_\textsc{r}]^\dagger  \Pi_q^\psi[f_\textsc{r}]=O
\end{align}
whenever $O\in \mathfrak{A}(\text{R})$ with R being in the in-region of supp$f$, i.e. not intersecting its causal future.  

Note that the suggested notion of effective operation, relies on covariance, i.e. freedom to split on any spacelike Cauchy slice, and on being able to take the limit that we sketched above. This is to guarantee that, in a sense, there are no superluminal processes within the support of $f$. Importantly, resolving Sorkin's problem does not rely on this limiting procedure, but on the fact that there is always a choice of hypersurface such that $f_-$ will be fully spacelike to A, and so B cannot enable a superuminal signal from A to C (see figure \ref{fig:sorkin_cont}). Recall that, even in the case when continuous additivity is not available, impossible operations are excluded by the Hammerstein property, as shown in Appendix \ref{app:hammerstein}.

\section{Discussion}

\subsection{Generality of the results}\label{subsec:causal_properties_kraus}
The convolution property of Kraus operators in Eq \eqref{eq:example_kraus_factorisation} is key in our anaylsis of causality and a motivation for effective Kraus operators. It is thus natural that we investigate when this condition holds in the local $S$-matrix formalism. In this section, we wish to understand what assumptions are required for identity \eqref{eq:example_kraus_factorisation} to hold more generally in the formalism of local $S$-matrices obeying continuous additivity
\begin{equation} \label{contadd}
    S[f]=S[f_++f_-]=S[f_+]S[f_-].
\end{equation}
We will assume that the $S$-matrix takes the form of a quantum controlled unitary. A quantum controlled unitary can be seen as first reading the value of the control system (in our case the probe) and then applying a unitary to the system (here the field) based on that control value (here $p$) 
\begin{equation} \label{qcontrol}
   S[f]=\int \d p \,U_p[f] \otimes \ket{p}\bra{p}. 
\end{equation}
This is a direct generalisation of controlled gates in quantum circuits. The implication of continuous additivity for Eq \eqref{qcontrol} is that 
\begin{equation} \label{contup}
   U_p[f_+]U_p[f_-]=U_p[f].
\end{equation}
This tells us that local $S$-matrices that implement quantum control factorise into matrices that factorise pointwise, which implies that
\begin{align} \label{quantumcontrol}
    \int \d s &\Pi^{\psi_+}_{q-s}[f_+]\Pi^{\psi_-}_{s}[f_-]\nonumber
    \\=&\int \d s\braket{q-s|S[f_+]|\psi_+}\braket{s|S[f_-]|\psi_-}\nonumber\\=\frac{1}{2\pi}&\int \d s \d p \d q U_p[f_+]U_q[f_-] \tilde{\psi}_+(p)\tilde{\psi}_-(q)e^{ipq}e^{is(p-q)}
    \nonumber
    \\=&\int \d p U_p[f_+]U_p[f_-] \tilde{\psi}_+(p)\tilde{\psi}_-(p)e^{ipq}\nonumber\\=&\int \d p U_p[f] \tilde{\psi}(p)=\Pi^\psi_q[f]
\end{align}
where the penultimate equality comes from the usual convolution theorem, and the final equality comes from the assumption of continuous additivity, Eq \eqref{contup}, of $U_p[f]$. It should be noted that quantum control is not an assumption that holds in general.

For Kraus families consisting of a single element, the convolution identity takes the form of a Weyl identity:
\begin{equation}
    e^{iH_+}e^{iH_-}=e^{iH}e^{i\theta},
\end{equation}
where $H^\pm, H$ are localised in the obvious sense. This is clearly satisfied when $H=a+\phi(f), a \in \mathbb{R}$, and fails for more general polynomials of fields: given a generic Cauchy slice $\Sigma$
\begin{equation}
    e^{i\phi(f)^2}\neq e^{iH_+}e^{iH_-}e^{i\theta}
\end{equation}
for $H^\pm$ localised in $\supp{f}\cap J^\pm(\Sigma)$. This is expected as generically only polynomials of first order in the fields lead to no-signalling unitary kicks. More details on this will be provided in \cite{simmons_brukner_in_prep}. 

To summarise, we have shown that for some forms of $S$-matrix, i.e. quantum controlled, the Kraus operators directly inherit the factorisation property in the form of a convolution type identity as in Eq \eqref{eq:example_kraus_factorisation}. Quantum controlled unitaries of this form are a very restricted set of evolutions, specifically ones where the Hamiltonian of the probe commutes with the interaction. For two fields coupled together, this is not a realistic assumption. While Eq \eqref{eq:example_kraus_factorisation} cannot be expected to hold in general, no-signalling will whenever $S$-matrices factorise. Further, Eq \eqref{almosteffective} will also hold generally, whenever the $S$-matrix factorises, as 
\begin{align}
    S[f]^\dagger \mathfrak{A}(\text{C})S[f]=&S[f_-]^\dagger S[f_+]^\dagger \mathfrak{A}(\text{C})S[f_+]S[f_-]\nonumber\\=&S[f_-]^\dagger \mathfrak{A}(\text{C})S[f_-]
\end{align}
implies $\Phi[f](C)=\Phi[f_-](C)$. Hence, while the convolution identity we used appears to be overly restrictive, the conclusions we made from it 1) no signalling and 2) the appearance of effective Kraus operators, hold far more generally. This implies that the conclusions we have made using the convolution identity are solid, and mostly independent of the $S$-matrix's form, relying only on it factorising.

\subsection{Measuring observables beyond \texorpdfstring{$\phi(f)$}{phi(f)}}\label{subsec:measuring_other_observables}

We have demonstrated the existence of a simple and solvable model capable of measuring $\phi(f)$ for any $f$ in a fully causal manner. This model is consistent with the findings of \cite{PhysRevD.105.025003, albertini2023ideal, oeckl2024spectraldecompositionfieldoperators}. Specifically, the Kraus operators are determined by a suitable $L^2$-normalized function of $\phi(f)$ (in the sense of spectral calculus), which, in this model, depends on the choice of the probe's state. Notably, $\phi(f)$ is a unique operator within the algebra: products of smeared fields generally do not produce non-signalling unitary kicks. Since the non-signalling property of a Gaussian measurement of $\phi(f)$ can be derived from the non-signalling nature of unitary kicks by $e^{i\phi(f)}$ \cite{PhysRevD.105.025003}, one might expect Gaussian (or more general $L^2$-normalized) measurements of powers of smeared fields to also be non-signalling. However, it has been shown that for a broad class of products of smeared fields, Gaussian measurements do exhibit signalling \cite{PhysRevD.105.025003, albertini2023ideal}.

One way to approach this is to note that for sufficiently temporally thin $f$ and in a lattice approximation, $\phi(f)$ can be approximated as
\begin{align}
    \phi(f)\sim A_1\otimes \openone \dots \otimes \openone + \openone \otimes A_2 \dots \otimes \openone \nonumber\\ + \openone \otimes \openone \dots  A_N.
\end{align}
A compact operator of this form can be measured in a non-signalling way \cite{PhysRevD.104.025012} with an update map that projects onto its eigenstate, i.e. a non-destructive measurement. However, it is well known in quantum information \cite{PhysRevLett.90.010402, PhysRevA.49.4331} that compact `non-local' operators cannot be measured in a way that does not erase all local information. By non-local we mean that they cannot be written in a way that is local with respect to the tensor product decomposition, e.g. $A\otimes B+A'\otimes B'\neq \tilde{A}\otimes \openone +\openone \otimes \tilde{B}$. Analogously in field theory
\begin{equation}\label{eq:local_observable}
    F[\phi]=\int \d \mathsf x_1 \dots \d \mathsf x_n F(\mathsf x_1,\dots \mathsf x_n)\phi(\mathsf x_1)\dots \phi(\mathsf x_n) 
\end{equation}
is local if 
\begin{equation}
    F(\mathsf x_1,\dots \mathsf x_n)=f(\mathsf x_1)\delta(\mathsf x_1-\mathsf x_2)\dots \delta(\mathsf x_1-\mathsf x_n)
\end{equation}
and non-local otherwise. It is clear that $\phi(f)^2$, which does not lead to a non-signalling unitary kick (see section \ref{subsec:sig_and_non}), will be non-local in this way even for temporally thin $f$, and so the measurement should be highly destructive. This was first noted in the context of Sorkin's impossible operations in \cite{Gisin2024towardsmeasurement}. The implication here is that the Kraus operators that measure $\phi(f)^2$, and other products of smeared fields, should not be simple spectral functions of the operator, unlike the case of $\phi(f)$. We can see this by noting that the Kraus operators being spectral functions implies that if $\ket{\psi}$ is an eigenstate of the observable, its measurement disturbance will be, or can be made, small. This cannot occur if all local information is lost.

This leads us to conclude that there are two distinct consequences for the causality of QFT measurements: 1) dynamical assumptions preclude the approximations used in the usual von Neumann derivation of sharp ideal measurements, 2) non-local operators must be measured in such a way that the selective state update is destructive \cite{PhysRevA.49.4331, PhysRevLett.90.010402}. While a projective measurement or even the measurement model we have studied here has Kraus operators that are simple spectral functions of the observable $\phi(f)$, and so the final state is expected to be closely related to the initial state, for non-local observables, we expect the selective version of any causal update rule will lead to a final state that is very different from the original, destroying the initial state. 

\section{Conclusions}

The impossibility of importing many standard measurement channels into QFT leaves a large gap in its foundations. Here we have studied the relationship between non-signalling channels and unitaries using the local $S$-matrix approach to QFT. Within this approach there are natural conditions, namely a hierarchy of factorisation conditions, which are sufficient for ruling out Sorkin's impossible operations. Further, we have studied the operational implications for the resulting channels, leading to causality conditions for the Kraus operators and the notion of effective Kraus operators.

The results in this paper hinged on combining quantum information notions of Stinespring and Kraus representations, and abstract notions of local QFT channels, which we introduced section \ref{sec:ops_and_measurements_QFT}. We reviewed examples of seemingly reasonable local channels that nevertheless lead to superluminal signalling, as typified by Sorkin's impossible operations. In particular, in section \ref{subsec:sig_and_non}, the fact that $e^{i\phi(f)}$ does not signal while $e^{i\phi(f)^2}$ does, leads us to consider the various notions of factorisation for local $S$-matrices in section \ref{subsec:rel_dynamics}. 

We demonstrated that local $S$-matrices satisfying continuous additivity or the Hammerstein factorisation property cannot exhibit superluminal signalling. Notably, the Hammerstein property, being a weaker condition, is sufficient to ensure no-signalling. Additionally, we analysed the properties of the $S$-matrix and the corresponding Kraus operators induced on the field in Section \ref{subsec:causality_for_example}. Specifically, we showed that if continuous additivity holds, the Kraus operators can be chosen to be localized within a subregion arbitrarily close to being entirely within the causal past of the region they update. We refer to these as effective Kraus operators. While this property is stronger than what is required to rule out impossible measurements, it naturally arises from the local $S$-matrix formalism. Consequently, it is expected to be a feature of many physically realisable channels.

 With this general framework, in section \ref{sec:example} we studied an explicit example of a factorisable $S$-matrix $S[f]$, given by a von Neumann measurement model for field amplitudes $\phi(f)$. This model is exactly solvable, leads to an unsharp measurement of $\phi(f)$ and allows us to analyse whether $\phi(f)$ can be measured causally. The application of the $S$-matrix framework to this simple model has interesting consequences. Specifically, we see a lower bound on the sharpness of a measurement of $\phi(f)$ given by $|\Delta_R(f,f)|^{\frac{1}{2}}$, and thus explicit dependence on the causal connectivity of the coupling region. To reach a sharp measurement, one must replace $f\mapsto f\lambda$, and take $\lambda\rightarrow 0$, which amounts to changing the physical coupling (see e.g. \cite{mandrysch2024quantumfieldmeasurementsfewsterverch}), and while also taking the probe state to be more sharp in $Q$. Similarly, we find that the model exhibits a generalisation of the convolution identity found in \cite{oeckl2024spectraldecompositionfieldoperators} as a no-signalling condition for Kraus operators. We've argued that this identity is a special case of a more general property of Kraus operators, and follows from the quantum controlled nature of the $S$-matrix in the simple model.

The connection between factorisation (continuous additivity and the Hammerstein property) and ruling out impossible measurement scenarios is not particularly surprising; however, it establishes a direct link between an operational notion of no-signalling and QFT concept of causality—specifically, the factorization of the $S$-matrix across any Cauchy surface. This contrasts with the no-signalling conditions proposed in \cite{sorkin1993impossible,PhysRevD.104.025012, PhysRevD.105.025003,albertini2023ideal, Fuksa}, which are global in nature. These conditions depend non-trivially on the entire measurement setup and do not necessarily reduce to a requirement that must hold locally within a specific region, independent of the broader scenario. Since the factorisation condition must hold locally, it provides a more robust and general framework for ensuring causality in QFT. Finally, since the stronger form of factorisation, continuous additivity, can be shown to imply the microcausality of the interaction Lagrangian, this restores the connection between microcausality and no-signaling, aligning well with the AQFT framework.

As an outlook, it is worth commenting on the potential connections to a causal version of Stinespring's theorem. While the Stinespring dilation theorem is a general result about unital completely positive channels between $C^*$-algebras and thus holds abstractly in QFT, the idea of a causal Stinespring theorem has been suggested in the literature \cite{PhysRevD.105.025003, fewster_asymptotic_2023}. As Jubb puts it, “It would be useful to construct an explicit dictionary between update maps and specific probe models, and determine if this is possible in general. The latter would be analogous to Stinespring's dilation theorem, but with the added restriction of locality and causality on the unitary.” In this paper, we have shown that the abstract maps discussed in \cite{PhysRevD.105.025003, albertini2023ideal, oeckl2024spectraldecompositionfieldoperators}, when interpreted as functionals of $\phi(f)$, can be dilated to an $S$-matrix $S[f]$ that obeys factorisation. We found that the Kraus operators derived from these maps exhibit general compositional properties directly tied to the factorisation of $S[f]$. This result has the flavour of a causal Stinespring theorem or, at least, provides hints of relativistic constraints on dilated channels. Since our analysis has focused on a limited class of local $S$-matrices when examining the properties of the induced Kraus operators, we cannot yet make general claims. Nevertheless, we believe this work offers valuable insights toward the development of a causal Stinespring theorem, something we will present general results on elsewhere \cite{simmons_brukner_in_prep}.

\begin{acknowledgments}
This research was funded in whole or in part by the Austrian Science Fund (FWF)
[10.55776/F71] and [10.55776/COE1]. For open access purposes, the authors have applied a CC BY public copyright
license to any author accepted manuscript version arising from this submission. This publication was made possible through the financial support of WOST (WithOutSpaceTime) grant from the
John Templeton Foundation.
The opinions expressed in this publication are those of the authors and do not necessarily reflect the views of the John Templeton Foundation. The authors are grateful to Robert Oeckl and Jos\'e de Ramon Rivera for detailed feedback during this work. We also thank Jan Mandrysch and Miguel Navascués for discussions on the connections between the pointer variable model and models of the FV framework. We are also grateful to Francisco Calderón and Doreen Fraser for insightful comments on earlier versions of this draft. Finally, the authors have benefited from the activities of COST Action CA23115:
Relativistic Quantum Information, funded  by COST (European Cooperation
in Science and Technology) and, in particular, from discussions at EMS-IAMP Spring School:
Symmetry and Measurement
in Quantum Field Theory at York, UK.
\end{acknowledgments}

\clearpage
\onecolumngrid

\makeatletter
\onecolumn@grid@setup
\let\set@footnotewidth\set@footnotewidth@one
\makeatother

\appendix

\section{All operations with the Hammerstein property are possible}\label{app:hammerstein}
In this section we show that operations induced by local $S$-matrices obeying the first three conditions of section \ref{subsubsec:properties_of_smatrices} and the Hammerstein property \cite{rejzner2019localitycausalityperturbativeaqft},
\begin{equation}
    S[f_1+f_2+f_3]=S[f_1+f_2]S[f_2]^{-1}S[f_2+f_3],
\end{equation}
are not impossible operations. In order to do so, we consider a compact region B, with $\overline{\supp f}\subset$B. Then, for any two regions A and C obeying the Sorkin ordering, i.e. A$\prec$ B, and B$\prec$ C but A and C spacelike, we can compare the following expectations
\begin{equation}
    \text{Tr}(\rho \mathcal{E}_\textsc{a}(S[f]^\dagger C S[f]))-\text{Tr}(\rho S[f]^\dagger CS[f])=\delta,
\end{equation}
where $\mathcal{E}_\textsc{a}$ is local to A and $C\in \mathfrak{A}(\text{C})$. If $\delta=0$ for all $\rho, C$ and all Sorkin scenarios, then $S[f]$ is non-signalling. 
\begin{figure}[ht]
    \centering
    \includegraphics[width=0.55\linewidth]{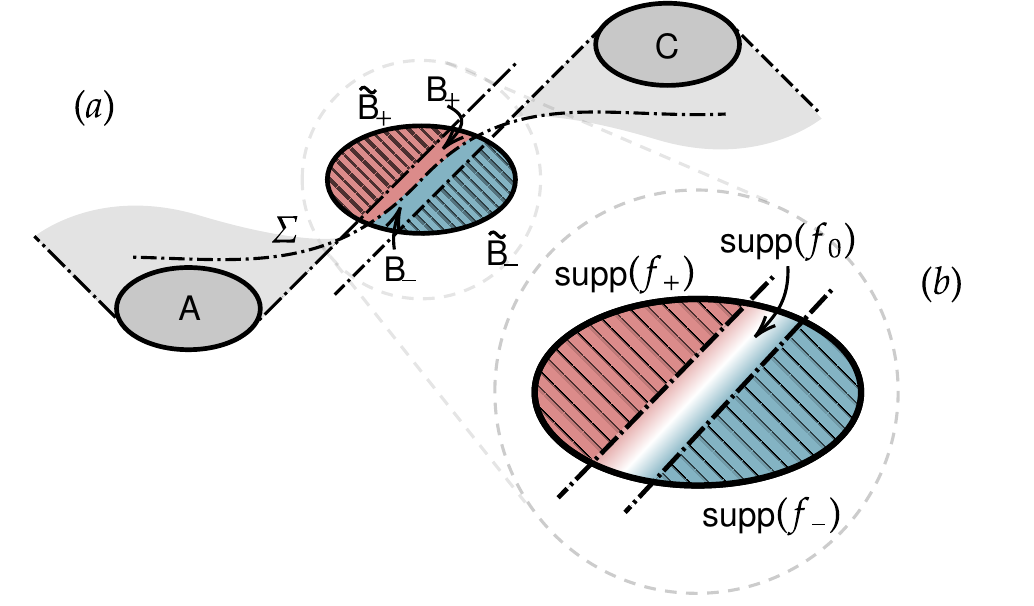}
    \caption{Location of the regions used in the proof of no Sorkin impossible operations (subfigure $(a)$), with a zoomed in view of the intermediate region (subfigure $(b)$). Subfigure $(a)$: The region B is the coloured intermediate region, while A and C are strictly spacelike separated from each one. Thus $\Sigma$ can be chosen such that A is in its past, and C in its future. The region $\text{B}_+$, which contains $\supp{f_+}$ is defined as the whole red region, while $\tilde{\text{B}}_+$, where $f_+=f$, is the hatched red region. Likewise, $\text{B}_-$, where $\supp{f_-}$ is contained, is the whole blue region, while $\tilde{\text{B}}_-$, where $f_-=f$, is the hatched blue region. The inclusions $\tilde{\text{B}}_\pm\subsetneq\text{B}_\pm$ are obvious, as are the causal relations. Note that the support of $f_0$ is localised in $\text{B}\backslash \tilde{\text{B}}_+\cup \tilde{\text{B}}_-$, i.e. the blue and red unhatched region. Subfigure $(b)$: The supports of $f_\pm,f_0$ are shown and coloured in red, blue and white respectively, with gradients representing overlapping supports. Note that there is no overlap in supports between $f_+$ and $f_-$. We can see that $f_++f_0$ has support only to the future of the hatched blue region, indicating that it is spacelike to C. Likewise, the support of $f_-+f_0$ is strictly in the past of the hatched red region, and thus spacelike to A. }
    \label{fig:hammerstein}
\end{figure}
By definition, there exists a Cauchy surface $\Sigma$ such that $J^+(\Sigma)\cap \text{B}=\text{B}_+$ is spacelike to C and $J^-(\Sigma)\cap \text{B}=\text{B}_-$ is spacelike to A. Also consider the following two regions $\tilde{\text{B}}_+=J^+(\text{A})\cap \text{B}$ and $\tilde{\text{B}}_-=J^-(\text{C})\cap \text{B})$. Note that $\tilde{\text{B}}_\pm \subsetneq \text{B}_\pm$, see figure\ref{fig:hammerstein}. Then we define the following (non-unique) continuous functions $f_\pm$ 
\begin{equation}
    \;\; f_\pm|_{\text{B}_\mp}=0, \;\; f_\pm|_{\tilde{\text{B}}_\pm}=f, \;\; f_\pm|_\Sigma=0
\end{equation}
i.e. $f_\pm$ vanishes in the future (past) of $\Sigma$, and coincides with $f$ in the future of A (past of C). Finally, let 
\begin{equation}
    f_0=f-f_+-f_-.
\end{equation}
The support of $f_0$ is in $\text{B}\backslash \tilde{\text{B}}_+\cup \tilde{\text{B}}_-$ (strictly, we should make sure that $f_\pm$ coincides with $f$ also within some non-zero distance of the boundary of $\tilde{\text{B}}_\pm$, so that $\supp{f_0}$ is strictly spacelike to A and C), i.e. the unhatched blue and red region of figure\ref{fig:hammerstein}. Then we have that $f_0$ is spacelike to A and C (note that $f_\pm|_{\tilde{\text{B}}_\pm}=f$ implies $f_0|_{\tilde{\text{B}}_\pm}=0$). This implies that $S[f_+]$ and $S[f_++f_0]$ commute with $\mathfrak{A}(\text{C})$ for example. We finally have the ingredients to show that $\delta=0$.
\begin{align}
    \text{Tr}(\rho \mathcal{E}_\textsc{a}(S[f]^\dagger C S[f]))=&\text{Tr}(\rho \mathcal{E}_\textsc{a}(S[f_-+f_0]^\dagger S[f_0]S[f_++f_0]^\dagger C S[f_++f_0]S[f_0]^\dagger S[f_-+f_0]))\nonumber\\
    =&\text{Tr}(\rho \mathcal{E}_\textsc{a}(S[f_-+f_0]^\dagger S[f_0] CS[f_0]^\dagger S[f_-+f_0]))\nonumber\\=& \text{Tr}(\rho \mathcal{E}_\textsc{a}(S[f_-+f_0]^\dagger C  S[f_-+f_0]))\nonumber\\
    =& \text{Tr}(\rho(S[f_-+f_0]^\dagger C  S[f_-+f_0]))\nonumber\\
    =& \text{Tr}(\rho(S[f_-+f_0]^\dagger S[f_0]S[f_++f_0]^\dagger C S[f_++f_0]S[f_0]^\dagger S[f_-+f_0]))
    \nonumber 
    \\=&\text{Tr}(\rho (S[f]^\dagger C S[f])).
\end{align}
To reach the second and third lines, we have used that $S[f_++f_0]$ and $S[f_0]$ commute with $\mathfrak{A}(\text{C})$ respectively. To reach the fourth line, we have used that $S[f_-+f_0]^\dagger C S[f_-+f_0]$ is a product of operators that are localised spacelike to A, so $\mathcal{E}_\textsc{a}$ acts trivially by locality. Finally, we have used unitarity of the $S$-matrices and cyclicity of the trace to reach the final line. This implies $\delta=0$, and since we have not made any assumptions about the regions, or $\rho, C$, we can conclude that $S[f]$ is possible if it satisfies the Hammerstein property. Specifically, we can see that $\delta=0$ for all conditions implies $[S[f]^\dagger \mathfrak{A}(\text{C}) S[f], \mathfrak{A}(\text{A})]=0$ if $[\mathfrak{A}(\text{C}), \mathfrak{A}(\text{A})]=0$, which implies that it is a non-signalling evolution \cite{PhysRevD.104.025012}. 


\section{Smearing with discontinuous test functions }\label{app:temporal_switch}
In Appendix \ref{appendix:additivity} we will show that the $S$-matrix 
\begin{equation}
    S[f]=\int \text{d}p \;e^{-p^2\frac{i}{2}\Delta_\textsc{R}(f_\pm, f_\pm)} e^{-ip\phi(f_\pm)},
\end{equation}
derived in Section \ref{sec:example} obeys the continuous additivity condition. For this condition to even make sense, we need to show that after temporally splitting a generic test function $f=f_-+f_+=\chi_-f+\chi_+f$, $S[f_\pm]$ is well defined, where we have viewed the split as modulated by a pair of switching functions $0\leq \chi_\pm\leq 1$ with $\chi^++\chi^-=1$, which in the case of Appendix \ref{appendix:additivity} are discontinuous. In this section we sketch a proof that $\Delta_\textsc{R}(f_\pm, f_\pm),\; e^{i\phi(f_\pm)}$ are well defined for any switching functions $\chi^\pm$ and $f\in C^\infty_0(\mathcal M)$ with $\supp{f}\subset \text{K}$ for some compact $\text{K}$, such that $f\chi^\pm\in L^2_\text{K}(\mathcal M)$, where we define $L^2_\text{K}(\mathcal M)$ to be the space of $L^2$ integrable functions on $\mathcal M$, supported in $\text{K}$. We recall $\Delta:C^\infty_0(\mathcal M)\rightarrow \text{Sol}(\mathcal M)$, and $K:\text{Sol}(\mathcal M)\rightarrow \mathcal H_1$, where $\Delta$ is the Pauli Jordan propagator and $K$ is the map to the one particle Hilbert space  $\mathcal H_1$, and densely embeds the space of compact solutions, see \cite{wald1994quantum} for more details. 

Since the switching function $\chi_\pm$, $e^{i\phi(x)}$, and $\Delta_\textsc{R}$ are distributional, we must prove that their multiplications are well defined. For the Weyl operator, we use the fact that each $f_\pm$ corresponds to a vector in the one particle Hilbert space. For the retarded propagator, we show that its diagonal is a well defined bi-functional such that for any regularised sequence $C^\infty_0(\mathcal M)\ni f_\pm^n\rightarrow f_\pm$ the limit is the same.  

We define $L^2_\text{K}(\mathcal M)\ni f_\pm^n=\chi^nf_\pm\rightarrow f_\pm\in L^2_\text{K}(\mathcal M)$, where $\chi^n_\pm$ is a sequence of smooth switchings that converge to $\chi_\pm$ in $L^2(\mathcal M)$. We assume that $K\Delta:C^\infty_\text{K}(\mathcal M)\rightarrow \mathcal H_1$ extends to a unique continuous map from $L^2_\text{K}(\mathcal M)$. Equivalently, for some constant $C_\text{K}$,
\begin{equation}\label{eq:bound_hilbert}
    ||K\Delta(f_\pm^n-f_\pm)||_{\mathcal H_1}\leq C_\text{K}||f^n_\pm-f_\pm||_{L^2_\text{K}(\mathcal M)}
\end{equation}
and so $K\Delta f_\pm^n=\ket{\psi_n}\in \mathcal H_1$ is a Cauchy sequence for any Cauchy $f_\pm^n$. Finally, we can conclude that $W(f_\pm^n)=e^{i\phi(f_\pm^n)}\rightarrow W(f_\pm)=e^{i\phi(f_\pm)}$ in strong operator topology. 

We provide a sketch for the bound in the case of $m>0$ in Minkowski spacetime, with the assumption that it also holds more generally. Since $||K\Delta f||_{\mathcal H_1}^2=\braket{\Omega|\phi(f)^2\Omega}=W(f,f)$, we have
\begin{equation}
   ||K\Delta f||_{\mathcal H_1}^2=\int \frac{d^3 \vec k}{(2\pi)^32E(\vec k)} |\tilde f(E(\vec k), \vec k)|^2\leq \int \frac{d^3 \vec k}{(2\pi)^32E(\vec k)} T_\text{K}\int_I dt |\tilde f(t,  \vec k)|^2
\end{equation}
where $\tilde f(t,  \vec k)$ is the spacial Fourier transform of $f$, $E(\vec{k})=m^2+\vec k^2$, and $T_\text{K}$ is the temporal length of $\text{K}$. Since $1/E(\vec k)\leq 1/m$, we have 
\begin{equation}
   ||K\Delta f||_{\mathcal H_1}^2\leq \frac{T_\text{K}}{(2\pi)^32m}\int dtd^3 \vec k |\tilde f(t,  \vec k)|^2\leq C_\text{K}||f||_{L^2_\text{K}(\mathcal M)}.
\end{equation}

We now turn our attention to the retarded propagator. We wish to define the following quadratic functional $D_\textsc{R}^\pm(f):=D_\textsc{R}\chi_\pm(f):=\Delta_\textsc{R}(f_\pm,f_\pm)$. We use \cite[Theorem 6.5.3]{Duistermaat_Hormander_1972} \footnote{We use the notation from Fewster's presentation \cite[Corollary 3.2, Appendix A.2]{2023}, where $H^s_\text{K}$ is the Sobolev space defined by $C^\infty(\text{K})$ with the norm $||f||_{H^s}=||((-\Delta_\text{Laplacian}+I)^{s/2}f||_{L^2}$. Note that for $s=0$, this is simply the $L^2$ norm. Strictly, in Fewster's notation, \cite{Duistermaat_Hormander_1972} shows that $\Delta_\textsc{R}$ extends to $\Delta_\textsc{R}:H_0^0\rightarrow H^1_\text{loc}$, which are respectively the Soblev spaces $H_0^1=\bigcup_{K}H_\text{K}^1$ and $H^s_0=\{u \text{ distribution }| \chi u\in H^s_0
\text{ for all } \chi\in C^\infty_0\}$. Since $H^0_0=\bigcup_\text{K}L^2_\text{K}$, and $H^1_\text{loc}\subset H^0_\text{loc}=:L^2_\text{loc}(\mathcal M)$, we get our extension.} to get
\begin{equation}\label{eq:extension}
    \Delta_\textsc{R}:L^2_\text{K}(\mathcal M)\rightarrow L^2_\text{loc}(\mathcal M),
\end{equation}
where $L^2_\text{loc}(\mathcal M)$ is the space of functions that are square integrable after being smeared with $C^\infty_0(\mathcal M)$ test function. 
It is clear that $f_\pm =\chi_\pm f\in L^2_\text{K}(\mathcal M)$, and so the restriction
\begin{equation}
    f_\pm \mapsto \Delta_\textsc{R}f_\pm|_\text{K}
\end{equation}
is continuous. The bi-linear form
\begin{equation}
    D(l,m):=\int_\text{K}\text{d}\mathsf x\text{d}\mathsf y \; l(\mathsf x)\Delta_\textsc{R}(\mathsf x-\mathsf y)m(\mathsf y)=\braket{l, Dm}_\text{K}
\end{equation}
is then also continuous. Letting
\begin{equation}
    D_\textsc{R}^\pm(f):=\braket{f_\pm, \Delta_\textsc{R} f_\pm}_{L^2_\text{K}}=D(f_\pm, f_\pm),
\end{equation}
the continuity implies that it is independent of regularisation scheme for $\chi_\pm$, and so it is well defined. Note the temporal nature of the switching played no role in either of the well defined-ness arguments. We conclude that for any switching $\chi^\pm\in L^2_\text{loc}(\mathcal M)$, continuous or otherwise, 
\begin{equation}
    S_p[f]=e^{-p^2\frac{i}{2}\Delta_\textsc{R}(f_\pm, f_\pm)} e^{-ip\phi(f_\pm)}
\end{equation}
is well defined.
\vspace{-1cm}
\section{Continuous additivity of the local scattering maps in the linear pointer coupling} \label{appendix:additivity}
We wish to show that \eqref{Scontroled} satisfies
\begin{equation}\label{eq:add_2}
    S[f]=S[f_+]S[f_-],
\end{equation}
where $f\in C_0(\mathcal{M})$, $f_+=\Theta f, f_-=(1-\Theta)f$ where $\Theta(x)=\Theta(t)$ for some foliation $\mathcal{M}\cong\mathbb{R}_t\times \mathbb{R}^3$. Note that the right hand side of Eq \eqref{eq:add_2} makes sense due to Appendix \ref{app:temporal_switch}.

In order to see Eq \eqref{eq:add_2}, we use the definitions $[\phi(f), \phi(g)]= i \Delta (f,g)$ where 
\begin{equation}
   \Delta (f,g)= \Delta_{\textsc{a}} (f,g)-\Delta_{\textsc{r}} (f,g)  
\end{equation}
where $\Delta_{\textsc{a},\textsc{r}}$ the advanced and the retarded Green's functions respectively. 
Note that 
\begin{equation}
    \Delta(f_+\!,f_-)=-\Delta(f_-\!,f_+)=-\Delta_{\textsc{r}}(f_-\!,f_+) \label{id1}
\end{equation}
and 
\begin{equation}
\Delta_{\textsc{r}}(f_{+}\mkern-5mu+\mkern-3mu f_{-}\mkern-5mu,f_{+}\mkern-5mu+\mkern-3mu f_{-})=\Delta_{\textsc{r}}(f_+\!,f_+)+\Delta_{\textsc{r}}(f_-\!,f_-) +\Delta_{\textsc{r}}(f_-\!,f_+) \label{id2}
\end{equation}
since $\Delta_{\textsc{a}}(f_-\!,f_+)=0$. Then, 
\begin{align}
    S[f_+]S[f_-]&=\int \d p\,\d q\, e^{-ip\phi(f_+)}e^{-iq\phi(f_-)}\otimes\ket{p}\braket{p|q}\bra{q}e^{-ip^2\Delta_{\textsc{r}}(f_+\!,f_+)/2-iq^2\Delta_{\textsc{r}}(f_-\!,f_-)/2} \nonumber \\
&=\int \d p\, e^{-ip\phi(f_+\!+f_-)}e^{-ip^2\Delta(f_+\!,f_-)/2}\otimes\ket{p}\bra{p}e^{-ip^2(\Delta_{\textsc{r}}(f_+\!,f_+)+\Delta_{\textsc{r}}(f_-\!,f_-))/2} \nonumber\\
&=\int \d p\, e^{-ip\phi(f_+\!+f_-)}e^{-ip^2\Delta_{\textsc{r}}(f_-\!,f_+)/2}\otimes\ket{p}\bra{p}e^{-ip^2(\Delta_{\textsc{r}}(f_{+}\!+ f_{-}\!,f_{+}\!+f_{-})-\Delta_{\textsc{r}}(f_-\!,f_+))/2} \nonumber\\
&=\int \d p\, e^{-ip\phi(f_+\!+f_-)}\otimes\ket{p}\bra{p}e^{-ip^2\Delta_{\textsc{r}}(f_{+}\!+f_{-}\!,f_{+}\!+f_{-})/2}\nonumber\\
&=S[f_{+}\mkern-5mu+\mkern-3mu f_{-}]
\end{align}
where the 2nd line comes from BCH 
\begin{equation} \label{BCH}
e^{i\phi(f)}=e^{i\phi(f_+)}e^{i\phi(f_-)}e^{-i\Delta(f_+,f_-)/2},
\end{equation}
and the 3rd from the propagator identities \eqref{id1} and \eqref{id2} above. 
\section{Non-signalling convolution property of the Kraus operators in the linear pointer coupling} \label{causal convolution}
Suppose $\psi$ is a normalised $L^2(\mathbb{R})$ function, then the Kraus operators for a field coupled to a probe in state $\braket{x|\psi}=\psi(x)$, using the decomposition \eqref{Scontroled}, are given by
\begin{align}
    \Pi^\psi_q[f]&=\langle x| S[f]|\psi \rangle \nonumber \\
    &=\frac{1}{\sqrt{2\pi}}\int \d p\, \tilde{\psi}(p)\exp{\left(-\frac{i}{2}p^2\Delta_{\textsc{r}}(f,f)+ip(q-\phi(f))\right)},
\end{align}
where $\tilde{\psi}$ is the Fourier transform of $\psi$. This simple model satisfies the following convolution property, which was been proven to be relevant for ensuring no-signalling \cite{oeckl2024spectraldecompositionfieldoperators}
\begin{equation}
         \Pi^{ \psi}_q[f]=\int \d s\, \Pi^{\psi_+}_{q-s}[f_+]\Pi^{\psi_-}_{s}[f_-],
\end{equation}
where $f_{\pm}$ are defined in the previous subsection, and $\psi_{\pm}$ are an arbitrary decomposition of the initial state of the probe $\psi$.
To see this, we start from  
\begin{align}
    \int \d s\, \Pi^{\psi_+}_{q-s}[f_+]\Pi^{\psi_-}_{s}[f_-]=\frac{1}{2\pi}\!\int \d s\, \d p\, \d q\, \widetilde{\psi_+}(p)\widetilde{\psi_-}(q)& \exp{\left(\!-\frac{i}{2}p^2\Delta_{\textsc{r}}(f_+,f_+)+ip(q\!-\!s\!-\!\phi(f_+))\right)} \nonumber\\
    \times&  \exp{\left(-\frac{i}{2}q^2\Delta_{\textsc{r}}(f_-,f_-)+iq(s-\phi(f_-))\right)}
\end{align}
Performing the $s$ and then $q$ integrals and using the BCH \eqref{BCH}
\begin{align}
\int \d s\, &\Pi^{\psi_+}_{q-s}[f_+]\Pi^{\psi_-}_{s}[f_-]\nonumber \\
&=\int \d p\,\widetilde{\psi_+}(p)\widetilde{\psi_-}(p)\exp{\left(-\frac{i}{2}p^2(\Delta_{\textsc{r}}(f_+,f_+)+\Delta_{\textsc{r}}(f_-,f_-))+ipq\right)} \exp{\left(-ip\phi[f_+]\right)}\exp{\left(-ip\phi[f_-]\right)}
\nonumber \\
    &=\int \d p\,\widetilde{\psi_+}(p)\widetilde{\psi_-}(p)\exp{\left(-\frac{i}{2}p^2(\Delta_{\textsc{r}}(f_{+}\!+f_{-},f_{+}\!+f_{-}))+ipq\right)}  \exp{\left(-ip\phi(f_{+}\!+f_{-})\right)} \nonumber \\
   &=\frac{1}{\sqrt{2\pi}}\int \d p\,\widetilde{\psi_+\star\psi_-}(p)\exp{\left(-\frac{i}{2}p^2(\Delta_{\textsc{r}}(f_{+}\!+f_{-},f_{+}\!+f_{-}))\right)} \exp{\left(ip(q-\phi(f_{+}\!+f_{-})\right)} \nonumber \\
    &=\Pi^{\psi_+\star\psi_-}_q[f_{+}\!+f_{-}]
\end{align}
where
\begin{equation}
    \widetilde{\psi_+\star\psi_-}(p)=\frac{1}{2\pi}\widetilde{\psi}(p).
\end{equation}

\section{A useful identity for the Kraus operators}\label{app:identity}
Another identity of the Kraus operators that we need for our derivation of non-signalling is the following 
\begin{equation}
    \int \d q \Pi^\psi_{q-s}[f]^\dagger\Pi^{\psi}_{q-s'}[f]= F_{\psi} (s-s')\openone_{\text{field}}.
\end{equation}
To see this, 
\begin{align}
 \int &\d q \Pi^\psi_{q-s}[f]^\dagger\Pi^{\psi}_{q-s'}[f]\nonumber \\
 &=\frac{1}{2\pi}\int \d q \d p \d k\, \widetilde{\psi}(p)^* \widetilde{\psi}(q)
 \exp{\left(\frac{i}{2}p^2\Delta_{\textsc{r}}(f,f)-ip(q-s-\phi(f))\right)}\exp{\left(-\frac{i}{2}k^2\Delta_{\textsc{r}}(f,f)+ik(q-s'-\phi(f))\right)} \nonumber \\
    &=\frac{1}{2\pi}\int \d q \d p \d k\, \widetilde{\psi}(p)^*\widetilde{\psi}(q)e^{-i(p-k)q}e^{ips-iks'} \exp{\left(\frac{i}{2}(p^2-k^2)\Delta_{\textsc{r}}(f,f)+i(p-k)\phi(f))\right)}\nonumber\\
    &=\int \d p \, |\widetilde{\psi}(p)|^2e^{ip(s-s')} \openone_\text{field}\nonumber := F_\psi(s-s') \openone_\text{field}.
\end{align}
Note that for $s=s'$ and for $\langle \psi| \psi \rangle =1$ this reduces to the usual trace preserving condition for Kraus operators, 
\begin{equation}
    \int \d q \Pi^\psi_q[f]^\dagger\Pi^{\psi}_q[f]=\openone_\text{field}.
\end{equation}

\section{`Half' an operation is still an operation}\label{app:effective}
Here we show the equality in Eq \eqref{effectiveoperation}. Consider $C$ local to a region that is spacelike to the support of $f_+$. Undoing the last step of \eqref{eq:effective}, what we want to show is that
\begin{equation}
    \int \d q \d s\d s'\text{Tr}\left(C\Pi^{\psi_-}_{s'}[f_-]\rho\Pi^{\psi_-}_{s}[f_-]^\dagger\left(  \Pi^{\psi_+}_{q-s}[f_+]^\dagger \Pi^{\psi_+}_{q-s'}[f_+]\right)\right)= \int \d q \text{Tr}\left(\Pi_q^\psi[f_-]^\dagger C\Pi_q^\psi[f_-]\rho \right)
\end{equation}

In order to do so, we expand the left hand side as follows
\begin{align}
\int \d q \d s\d s'\text{Tr}\left(C\Pi^{\psi_-}_{s'}[f_-]\rho\Pi^{\psi_-}_{s}[f_-]^\dagger\left(  \Pi^{\psi_+}_{q-s}[f_+]^\dagger \Pi^{\psi_+}_{q-s'}[f_+]\right)\right)=
    \frac{1}{(2\pi)^2}\int \d q\d s\d s'\d p\d n\d k\d m \widetilde{\psi_+}^*(p)\widetilde{\psi_-}^*( n)\widetilde{\psi_+}(k)\widetilde{\psi_-}(m)\nonumber\\\times \text{exp}\left(\frac{i}{2}(\Delta_{\textsc{r}}(f_+,f_+)(p^2-k^2)+\Delta_{\textsc{r}}(f_-,f_-)(n^2-m^2)-ip(q-s)-iqs+ik(q-s')+ims'\right)\nonumber\\ \times \text{Tr}\left(Ce^{-im\phi[f_-]}\rho e^{in\phi[f_-]} e^{-i\phi[f_+](p-k)}\right)
\end{align}
Performing the $s,s'$ integrals, we get,
\begin{align}
    \int \d q \d s\d s'\text{Tr}\left(C\Pi^{\psi_-}_{s'}[f_-]\rho\Pi^{\psi_-}_{s}[f_-]^\dagger\left(  \Pi^{\psi_+}_{q-s}[f_+]^\dagger \Pi^{\psi_+}_{q-s'}[f_+]\right)\right)= \int \d q\d p \d k\widetilde{\psi_+}^*(p)\widetilde{\psi_-}^*(p)\widetilde{\psi_+}(k)\widetilde{\psi_-}(k) \nonumber\\
    \times \text{exp}\left(\frac{i}{2}(\Delta_{\textsc{r}}(f_+,f_+)+\Delta_{\textsc{r}}(f_-,f_-))(p^2-k^2)-iq(p-k)\right)
\text{Tr}\left (Ce^{-ik\phi[f_-]}\rho e^{ip\phi[f_-]} e^{-i\phi[f_+](p-k)}\right)\nonumber\\
=\int\d p |\widetilde{\psi}(p)|\text{Tr}\left (e^{ip\phi[f_-]}Ce^{-ip\phi[f_-]}\rho\right ),
\end{align}
where in the last step we performed the $x$ integral and then $k$ integral and used that $\widetilde{\psi_+\star\psi_-}=\frac{1}{2\pi}\widetilde{\psi}$.
Note that the Dirac delta $\delta(k-p)$ ensures that the factors including $\phi(f_+)$ disappear. Finally, to reintroduce the expression of the local Kraus operators (using the fact that the update map is not sensitive to any phase multiplying the Kraus operators) we perform  the following trick
\begin{align}
\int\d p |\widetilde{\psi}(p)|\text{Tr}\left (e^{ip\phi[f_-]}Ce^{-ip\phi[f_-]}\rho \right )
=\frac{1}{2\pi} \int &\d q \d p  \d k\widetilde{\psi}(p)^*\widetilde{\psi}(q)\text{exp}\left(\frac{i}{2}(p^2-k^2)\Delta_{\textsc{r}}(f_-,f_-)-iq(p-k)\right)\nonumber\\
&\times\text{Tr}\left(e^{ip\phi[f_-]}Ce^{-ik\phi[f_-]}\rho\right)
=\int \d q \text{Tr}\left(\Pi_q^\psi[f_-]^\dagger C\Pi_q^\psi[f_-]\rho\right)
\end{align}
as claimed.

\bibliography{references.bib}
\end{document}